\documentclass[%
 reprint,onecolumn,
 amsmath,amssymb,
 aps, showkeys
]{revtex4-2}
\usepackage{amsfonts}
\usepackage{graphicx}
\usepackage{dcolumn}
\usepackage{bm}
\usepackage{stmaryrd}
\usepackage{color}
\usepackage{cancel}
\usepackage{mathtools}
\usepackage{upgreek}
\graphicspath{{./figures/}}

\newcommand{\blue}[1]{\textcolor{black}{#1}}

\newcommand{\beq}{\begin{equation}}
\newcommand{\eeq}{\end{equation}}
\newcommand{\pfr}[2]{\ensuremath{\frac{\partial #1}{\partial #2}}}
\newcommand{\pfi}[2]{\ensuremath{{\partial #1}/{\partial #2}}}
\newcommand{\ep}{\varepsilon}

\newcommand\Pra{\mbox{\textit{Pr}}}
\newcommand\Lew{\mbox{\textit{Le}}}

\newcommand{\vect}[1]{\mathbf{#1}}

\newcommand{\be}{\begin{equation}}
\newcommand{\ee}{\end{equation}}
\newcommand{\bea}{\begin{eqnarray}}
\newcommand{\eea}{\end{eqnarray}}
\newcommand{\beas}{\begin{eqnarray*}}
\newcommand{\eeas}{\end{eqnarray*}}

\def\ep{\epsilon}

\def\nn{{\bf n}}

\newcommand{\UU}{{\bf U}}
\def\uu{{\bf u}}
\def\vv{{\bf v}}

\begin{document}

\preprint{APS/123-QED}

\title{Hydrodynamic instabilities of propagating interfaces under Darcy's law}
\author{Joel Daou} 
 \email{joel.daou@manchester.ac.uk}
\author{Prabakaran Rajamanickam}\email{Present address: Department of Mathematics \& Statistics, University of Strathclyde, Glasgow G1 1XQ, UK}  
\affiliation{Department of Mathematics, University of Manchester, Manchester M13 9PL, UK}

%

%

\date{\today}

\begin{abstract}
The hydrodynamic instabilities of propagating  interfaces in Hele-Shaw channels or porous media under the influence of an imposed flow and gravitational acceleration are investigated within the framework of Darcy's law.  The stability analysis pertains to an interface between two fluids with different densities, viscosities, and permeabilities, which can be susceptible to Darrieus--Landau, Saffman--Taylor, and Rayleigh--Taylor instabilities. A theoretical analysis, treating the interface as a hydrodynamic discontinuity, yields a simple dispersion relation between  the perturbation growth rate $s$ and  its wavenumber $k$  in the form $s=(ak - bk^2)/(1+ck)$, where  $a$, $b$ and $c$ are constants determined by problem parameters. The constant $a$ characterises all three hydrodynamic instabilities, which are long-wave in nature. In contrast, $b$ and $c$, which characterize the influences of local curvature and flow strain on interface propagation speed, typically provide stabilisation at short wavelengths comparable  to interface's diffusive thickness. The theoretical findings for Darcy's law are compared with a generalisation of the classical work by Joulin \& Sivashinsky, which is based on an Euler--Darcy model. The comparison provides a conceptual bridge between predictions based on Darcy's law and those on Euler's equation and offers valuable insights into the role of confinement on interface instabilities in Hele-Shaw channels.  Numerical analyses of the instabilities are carried out for premixed flames using a simplified chemistry model and Darcy's law. The numerical results corroborate with the explicit formula with a reasonable accuracy. Time-dependent numerical simulations of unstable premixed flames are carried out to gain insights into the nonlinear development of these instabilities. The findings offer potential strategies for control of interface instabilities in Hele-Shaw channels or porous media. Special emphasis is given to the critical role played by the imposed flow in destabilising or stabilising the interface, depending on its direction relative to  the interface propagation.
\end{abstract}

\maketitle


\section{Introduction}
\label{sec:intro}

Premixed flames have long been known to exhibit two familiar hydrodynamic instabilities: the Darrieus--Landau (DL) instability and the Rayleigh--Taylor (RT) instability. The DL instability~\cite{darrieus1938propagation,landau1944slow} is experienced by a premixed flame, or any interface, propagating towards a denser fluid, while the RT instability~\cite{rayleigh1882investigation,taylor1950instability} occurs when gravity points away from the denser fluid. In addition to these instabilities, flames propagating in narrow geometries such as a Hele-Shaw channel or in a porous medium are prone to another hydrodynamic instability, namely the Saffman--Taylor (ST) instability.  First described by Saffman and Taylor \cite{saffman1958penetration} for a material liquid interface, the ST instability is encountered when a less viscous fluid displaces a more viscous fluid.

The role of Saffman--Taylor instability for a propagating interface such as a premixed flame   was first studied by Joulin and Sivashinsky~\cite{joulin1994influence}. Their analysis,
as well as subsequent ones in combustion theory, e.g.~\cite{kang2003computational,aldredge2004saffman,miroshnichenko2020hydrodynamic},  are not based on Darcy's law as in  Saffman--Taylor's original analysis, but use the so-called Euler--Darcy equation which combines Darcy's law with the inertial  terms of the Euler equation. Such models have been proposed based on heuristic arguments and approximation, see e.g.~\cite{gondret1997shear,ruyer2001inertial,plouraboue2002kelvin} and references therein.  They do not have however solid foundation based on a  consistent asymptotic derivation,  unlike the case of Darcy's law which can  be derived, in a Hele-Shaw channel for example, from the Navier--Stokes equation in the asymptotic limit of zero channel width. In other words, Darcy's law without the additional ad-hoc terms used in Euler-Darcy type equations, is the appropriate equation to use for an asymptotically correct leading-order description of the flow field in a narrow channel.   This observation is supported  by recent combustion investigations~\cite{fernandez2018analysis,martinez2019role,rajamanickam2024effect,fernandez2019impact} featuring Darcy's law. In this paper, we shall revisit the theoretical analysis developed in~\cite{joulin1994influence},  using the more appropriate Darcy's law as well as relaxing some of the restrictions adopted therein. As we shall demonstrate, this will lead to a more transparent description of the interface hydrodynamic instabilities in narrow channels or porous media.   
 
Another important motivation for this study is to clarify the  critical role played by an imposed flow in destabilising or stabilising the propagating interface, depending on the flow direction relative to the direction of propagation.  The  effect of the direction of a parallel flow  relative to that of flame propagation on the effective propagation speed has been addressed in~\cite{daou2001flame} both for wide and narrow channels. Its effect on hydrodynamic flame stability has been studied recently in~\cite{miroshnichenko2020hydrodynamic} based on the Euler--Darcy model, leading to a dispersion relation generalising that in \cite{joulin1994influence}. Outside the field of combustion, the stability of chemical fronts propagating in  Hele-Shaw cells or porous media  using Darcy's law has been addressed in several studies~\cite{vasquez1996chemical,de2001fingering,demuth2003chemical,d2007classification}. These have been conducted however for freely propagating interfaces, that is in the absence of an imposed flow, which has a crucial role when considering the Saffman--Taylor instability, as we shall demonstrate. Therefore, it is important  to distinguish between the stability of freely propagating interfaces, such as the flames investigated in the Hele-Shaw channel experiments of  Ronney, Almarcha and others~\cite{fernandez2018analysis,al2019darrieus,veiga2019experimental,veiga2020unexpected,dejoan2024effect}, from interfaces subjected to an imposed flow. Failure to appreciate this distinction may lead to erroneous interpretation of experimental and numerical findings when comparing with theoretical results.  It is worth pointing out for example that the stability dispersion relation derived by Joulin and Sivashinsky corresponds  in a Hele-Shaw channel to a single specific value  of the mean flow; the value which allows an undisturbed (depth-averaged) planar flame front to be stationary with respect to the channel walls. Strictly speaking, their results are not applicable to freely propagating flames, which are commonly considered in the literature~\cite{fernandez2019impact,chang2024experimental}. In this paper, we shall provide a simple dispersion relation combining the effect of the various hydrodynamic instabilities aforementioned and accounting for the presence of an imposed flow.   

The paper is structured as follows. A theoretical analysis is developed in Section~\ref{modelDarcy}, accounting for the interaction of the three hydrodynamic instabilities of a propagating interface in the presence of an imposed flow and gravitational acceleration.  In Section~\ref{modelEulerDarcy}, a brief  review of the results based on Euler--Darcy model is carried out for the purpose of comparison with  the pioneering work by Joulin and Sivashinsky \cite{joulin1994influence} and the   recent work by Miroshnichenko \textit{et. al.}~\cite{miroshnichenko2020hydrodynamic}. In principle,  the content of Sections~\ref{modelDarcy} and~\ref{modelEulerDarcy}   is of general validity and not restricted only to combustion applications. Section~\ref{numerical} presents a  numerical study based on Darcy's law focusing on premixed flame instabilities. Numerical simulations are reported, including eigenvalue computations and direct numerical simulations, which are used to validate and extend the analytical results.

 \begin{figure}
\centering
 \includegraphics[scale=0.75]{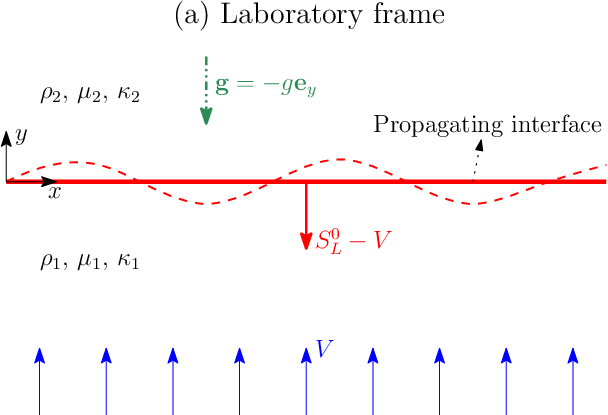}\hspace{1cm}
\includegraphics[scale=0.75]{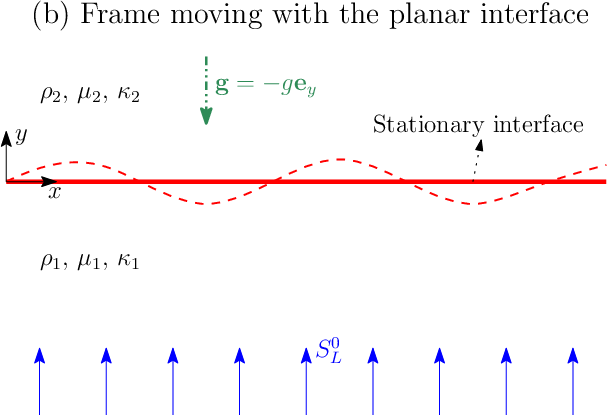}
\caption{(a) Schematic illustration, in the laboratory frame, of an unperturbed planar interface (solid line), which separates two media with different physical properties and propagates with a speed $S_L^0$ (with respect to upstream fluid 1); with respect to the lab, the speed of the planar interface is $S_L^0-V$. Dashed line indicates perturbed interface, which may propagate at a different speed due to kinematic and curvature-induced effects. (b) The corresponding illustration of the situation in the frame moving with the planar interface.} 
\label{fig:config}
\end{figure}

\section{Darrieus--Landau and Saffman--Taylor instabilities based on Darcy's law: Theoretical analysis} \label{modelDarcy}

\subsection{Formulation}

Consider an interface propagating with respect to the fluid below it (fluid 1) with local normal propagation speed $S_L$, as illustrated Fig.~\ref{fig:config}(a). The fluid is assumed to be flowing upwards from $y=-\infty$ with velocity $\uu = V \vect e_y$ in the laboratory frame. The density $\rho$, viscosity $\mu$, and permeability $\kappa$ are constant and equal to $\rho_1, \mu_1, \kappa_1$ in the lower fluid and  $\rho_2, \mu_2, \kappa_2$ in the upper.  For Hele-Shaw channels, $\kappa_1=\kappa_2=h^2/12$, where $h$ is the channel width. The flow field $\uu$ in the laboratory frame is assumed to follow Darcy's law, so that on each side of the interface
\begin{equation}
    \nabla\cdot\vect u = 0, \qquad \vect u = -\frac{\kappa}{\mu}  \nabla (p+\rho gy) \,, \label{darcylab}
\end{equation} 
from which it follows that the pressure field $p$ satisfies  Laplace's equation $\nabla^2 p=0$. 

In the unperturbed state, the interface is flat and  $S_L= S_L^0$, where $S_L^0$ is assumed to be a known constant.  In the laboratory frame, the interface therefore propagates downwards with constant speed $S_L^0-V$. It is convenient to choose a moving frame in which the unperturbed planar interface is stationary, assumed to be located at $y=0$, as illustrated in Fig.~\ref{fig:config}(b). In this frame the flow field is given by $\vect v=\vect u + (S_L^0-V)\vect e_y$  and we shall write $\vect v=(u,v)$ to define its components.  In the moving frame,   the governing equations  are given by
\begin{equation}
    \nabla\cdot\vect v = 0, \qquad \vect v = -\frac{\kappa}{\mu} \nabla (p+\rho gy) + (S_L^0-V)\vect e_y, \qquad \nabla^2p=0 \label{darcymove}
\end{equation}
on both sides of the interface,  and  $\vv = S_L^0 \,  \vect e_y$ as $ y \to - \infty$ as indicated in Fig.~\ref{fig:config}(b).

Let the (perturbed) interface be described by the equation $y=f(x,t)$ with  unit normal $\vect n$ pointing towards fluid $2$. By definition, the local propagation speed $S_L$ (with respect to fluid 1) is given by $S_L = \left. (\vv-\UU)\cdot \nn\right|_{y=f^-}$ where $\UU$ is the local interface velocity in the moving frame. In terms of $f$, we have
\begin{equation}
    \vect n = \frac{(-f_x,1)}{\left( 1+f_x^2 \right)^{\frac{1}{2}}} \,, \qquad   \qquad  \UU \cdot \nn = \frac{f_t}{\left( 1+f_x^2 \right)^{\frac{1}{2}}} \,.
\end{equation}
At the interface, the continuity of  
mass flux and  pressure and the definition of $S_L$ require
 \begin{equation}
    \llbracket \vect v\cdot\vect n \rrbracket = \left(\frac{\rho_1}{\rho_2}-1\right) S_L, \qquad \llbracket p\rrbracket =0, \qquad S_L = \left. \frac{v-u f_x -f_t}{\left( 1+f_x^2 \right)^{\frac{1}{2}}} \right |_{y=f^-} \,,  \label{interfacedarcy}
\end{equation}
with the notation  $\llbracket \varphi \rrbracket \equiv \varphi |_{y=f^+}-\varphi |_{y=f^-}$. The first condition follows from  the conservation of mass requirement
$\llbracket \rho(\vect v - \UU) \cdot \nn) \rrbracket =0$ and the definition of $S_L$. As for the second condition, this is obtained upon integrating,  across the interface, Darcy's equation (which is also valid   within the interface).

To close the problem formulation, it is important to define the local propagation speed $S_L$ of the curved interface appropriately.  Following Markstein~\cite{markstein1988experimental},   we shall assume that the deviation of $S_L$ from 
$S_L^0$ (the propagation speed of the unperturbed  planar interface)  has     a linear dependence  on the local interface curvature  of the form
\begin{equation}
    \frac{S_L}{S_L^0} = 1+\mathcal{L}\, \nabla\cdot \vect n   \label{SL}
\end{equation}
involving a Markstein length  $\mathcal{L}$.  Such dependence on curvature has not been taken into account in~\cite{joulin1994influence}, where it was assumed  that $S_L=S_L^0$ and where the physical meaning of the planar value $S_L^0$ is discussed, at least in the context of flame propagation in a Hele-Shaw channel. As emphasised in this reference, $S_L^0$ need not be the usual so-called laminar flame speed, but rather an effective (depth-averaged) value thereof which accounts for flame curvature along the Hele-Shaw channel wall's normal direction, as well as heat losses.  In principle, $S_L^0$ may also depend on the effective (depth-averaged) flow, $V$ in our notation, an assumption which was adopted in \cite{miroshnichenko2020hydrodynamic}.  The latter dependence is however negligible in the asymptotic limit  of zero channel width and hence will not be adopted in our study. 

More importantly, we note that the Markstein model~\eqref{SL}, although phenomenological in nature, is highly attractive due to its simplicity and its ability to realistically capture the effect of the interface curvature on its propagation speed.  Therefore, this model will be adopted in most part of this investigation.  It must be emphasised however that other  somewhat more complex models may be used, which are based on asymptotic analysis accounting for the inner  structure of the interface as well as local flow nonuniformities. This significant aspect is discussed in section~\ref{sec:ImprovedMark}.

\subsection{Linear stability of the planar interface}\label{sec:linearstabilitydarcy}

The basic solution   $(\vv,p,f)=\left(\overline \vv,  \overline{p},\overline{f}\right)$
corresponds to a stationary flat interface  with $\overline f =0$ and
\begin{align}
 &\overline \vv=S_L^0\vect e_y\,, \quad \quad  -\overline p= \rho_1 gy   + \frac{\mu_1}{\kappa_1}Vy \quad   \text{for} \quad  y<0  \,, \\
& \overline \vv =\frac{\rho_1}{\rho_2}S_L^0\vect e_y \,,  \quad  -\overline p=  \rho_2 gy + \frac{\mu_2}{\kappa_2}\left[V+\left(\frac{\rho_1}{\rho_2}-1\right)S_L^0\right]y \quad \text{for} \quad  y>0  \,,
\end{align}
which satisfies clearly the governing equations and auxiliary conditions. To the basic solution $\left(\overline \vv,  \overline{p},\overline{f}\right)$, we add small perturbations such that  $(\vv,p,f) = \left(\overline \vv,  \overline{p},\overline{f}\right) + (\vv',p',f')$ where the primed quantities satisfy the linearised governing equations
\[
 \nabla\cdot\vv' = 0 \,,  \quad    \vv'     = -\frac {\kappa}{\mu} \nabla p' \,,  \quad    \nabla^2 p' =0 \,.
\]
In terms of $p'$ and $f'$, the linearised interfacial conditions at $y=0$ can be written as   
\begin{equation} 
\left \llbracket \frac{\kappa}{\mu}  p'_y  \right \rrbracket  =  \left(\frac{\rho_1}{\rho_2}-1\right) S_L^0 \mathcal{L} f'_{xx} \,, \quad   \llbracket  p'\rrbracket =  \alpha f'  \,, \quad \left. f'_t = -\frac{\kappa_1}{\mu_1}  p'_y\right|_{y=0^-} + S_L^0 \mathcal{L} f'_{xx} \,,
\label{interface00}
\end{equation}
where
\begin{equation}
 \alpha = \left (  \frac{\mu_2}{\kappa_2} - \frac{\mu_1}{\kappa_1}  \right) V + (\rho_2-\rho_1) g +  \frac{\mu_2}{\kappa_2}\left (\frac{\rho_1}{\rho_2}-1 \right)S_L^0  \,.
\label{alpha}
\end{equation}
The final linear stability problem is given by $\nabla^2 p' =0$,  to be solved in the domains $y>0$ and $y<0$, subject to  the interfacial conditions~\eqref{interface00} at $y=0$  and the boundary condition $p'\to 0$   as $y \to \pm \infty$.

Since the stability problem does not depend explicitly  on $x$ and $t$, we   look for normal modes in the  form
\[
p'=\hat{p}(y) \exp \left( st+ikx \right) \quad {\rm and} \quad f'= C \exp \left( st+ikx \right)
\]
where $s$ is in general complex, $k$ real, and $C$ a constant. Then,  the equation  $\nabla^2 p' =0$ implies that $\hat{p}_{yy}- k^2 \hat{p} =0$, hence, using the the requirement $  p'=0$   as $y \to \pm \infty$, we have 
\[
\hat{p}= \left\{
  \begin{array}{lr}
   A e^{-k y}  &   \quad \text{for} \quad y>0 \,, \\
B e^{k y}  &    \quad \text{for} \quad y< 0 \,,
  \end{array}  
\right.
\]
where  $A$ and $B$ are constants and the wave number $k$ is assumed positive.   Upon imposing the three interfacial conditions~\eqref{interface00}, we obtain 
\begin{equation}  \label{matrixd}
     \begin{bmatrix}
           k\kappa_2/\mu_2 &    k\kappa_1/\mu_1 & -(\rho_1/\rho_2-1) S_L^0 \mathcal{L}  k^2 \\
        1 & -1 & -\alpha \\
        0 &    k\kappa_1/\mu_1 & s +  S_L^0 \mathcal{L} k^2
    \end{bmatrix}  \begin{bmatrix}
        A   \\
        B    \\
        C
    \end{bmatrix} =  {\bf 0}. 
\end{equation}
The solvability of this system of homogeneous equations yields the required dispersion relation 
\begin{equation}
    s = \frac{\alpha   k}{\mu_1/\kappa_1+\mu_2/\kappa_2} -   \frac{(\mu_2/\kappa_2)(\rho_1/\rho_2) +\mu_1/\kappa_1 }{\mu_1/\kappa_1+\mu_2/\kappa_2} S_L^0\mathcal L   k^2 \,, \label{dimdisp}
\end{equation}
where $\alpha$ is as given in~\eqref{alpha}.
A non-dimensional form of this dispersion relation is obtained by multiplying all terms by ${\delta_L^0}/ S_L^0$, where  
$\delta_L^0$ is the diffusive thickness of the flat interface,
and introducing the non-dimensional growth rate and wave number  $\tilde{s}= s \delta_L^0/S_L^0$ and $\tilde{k}= k \delta_L^0$, as well as the parameters
\begin{equation}
  r = \frac{\rho_1}{\rho_2}, \quad m = \frac{\mu_1/\kappa_1}{\mu_2/\kappa_2}, \quad  \mathcal V = \frac{V}{S_L^0}, \quad  G = \frac{\rho_2 g\kappa_2}{S_L^0 \mu_2} , \quad
   \mathcal M = \frac{\mathcal L}{\delta_L^0} \,.
   \label{nondim} 
\end{equation}
We thus obtain, dropping the tildes, the non-dimensional dispersion relation
\begin{equation}
   s = ak - b k^2,  \qquad \text {where} \quad  a= \frac{r-1}{1+m} + \frac{1-m}{1+m} \mathcal V + \frac{1-r}{1+m} G \quad \text{and} \quad  b = \frac{r+m}{1+m}\mathcal M. \label{nondimdisp}
\end{equation}

\subsection{Implication of the dispersion relation} \label{Sec:Implications}

\begin{figure}[h!]
\centering
\includegraphics[scale=0.45]{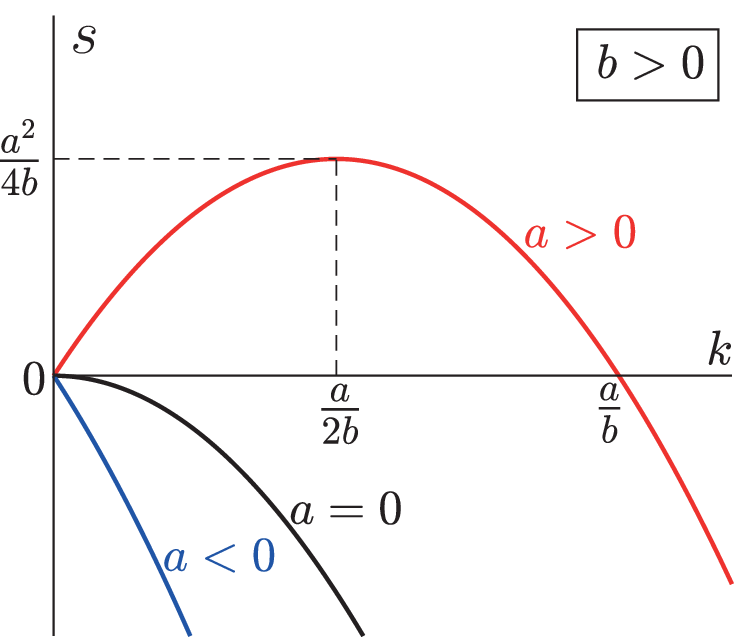}
\caption{Schematic illustration of the formula $s=ak-bk^2$ for $a>0$, $a=0$ and $a<0$, all with $b>0$. } \label{fig:sk}
\end{figure}

We are now able to discuss the stability of the interface implied by the dispersion relations~\eqref{dimdisp} or~\eqref{nondimdisp}. 
Since the ratios $r$ and $m$ are positive numbers, the stability is determined only by the two non-dimensional parameters $a$ and $\mathcal M$, or equivalently, by the dimensional parameters $\alpha$ and $\mathcal{L}$. 
Clearly, instability occurs when $a>0$ ($\alpha>0$)  or when $\mathcal{M}<0$ ($\mathcal{L}<0$).  
The second instability condition, $\mathcal M<0$,  
corresponds to a Turing-type instability known as the 
 diffusive-thermal instability in combustion. This instability, whose investigation requires  the analysis of the inner diffusive zone of the interface, will not be considered in this paper. Therefore, we shall assume henceforth that the Markstein length  $\mathcal{L}$ and the Markstein number $\mathcal{M}$ are positive. It follows that the necessary and sufficient condition for instability is that $a>0$ ($\alpha>0$), with the instability being hydrodynamic in nature. Specifically, the interface is prone to three hydrodynamic instabilities corresponding to the three terms in $a$ (or $\alpha$).   The first term  leads to the Darrieus--Landau instability when the density ratio $r>1$, the second term to the Saffman--Taylor instability when $(1-m)\mathcal V>0$, and the third term to the Rayleigh--Taylor instability when $(r-1)G<0$.  It is interesting to note that these three instabilities combine in a simple and transparent manner in the expression of $a$.   When $a>0$, the range of unstable modes is $k \in (0,b/a)$, with the most unstable mode having growth rate 
 $s=a^2/4b$ and wavenumber  $k= a/2b$.  A graphical illustration is provided in Fig.~\ref{fig:sk} for selected values of $a$. 

It is worth specialising the discussion to the case of flame propagation in Hele-Shaw channels which have been actively investigated recently~\cite{fernandez2018analysis,al2019darrieus,veiga2020unexpected,dejoan2024effect}. In this configuration, $\kappa_1=\kappa_2= h^2/12$, where $h$ is the channel width, and the dispersion relation~\eqref{dimdisp} takes the form  
\begin{equation}
     s = \left[ \frac{\mu_2(\rho_1-\rho_2) S_L^0}{\rho_2 (\mu_2+\mu_1)} +  \frac{ (\mu_2 - \mu_1) V}{\mu_2+\mu_1} +\frac{h^2 (\rho_2-\rho_1) g }{12 (\mu_2+\mu_1)}\right] k - \left(\frac{\rho_1 \mu_2 + \rho_2 \mu_1}{\rho_2 \mu_1+\rho_2 \mu_2} \right)S_L^0\mathcal L k^2 .
\end{equation}
Again the combined effect of the three hydrodynamic instabilities is encapsulated in the coefficient of $k$. 
 In premixed flames, since the density of the unburnt gas $\rho_1$ is larger than that of the burnt gas $\rho_2$, the first term in the square bracket is always destabilising, which is at the root of the DL instability. For the same reason, the third term is stabilising if $g>0$ and destabilising otherwise, which is at the root of the RT instability. Typically, this instability is comparatively weak in small channels due to the factor $h^2$ being then small. As for the second term, this is seen to be stabilising if $V<0$ and destabilising if $V>0$ since $\mu_2>\mu_1$ in flames given that the dynamic viscosity $\mu$ is an increasing function of temperature in gases. This term is at the root of the ST instability. When this term is large, the ST instability in dominant compared to the DL instability.
 For $V$ less than a critical value $V_c$, the flame is stable. This critical value is given, using~\eqref{nondim} with
 $m=\mu_1/\mu_2$  and $G=\rho_2 g h^2/12 S_L^0 \mu_2$ in this case, by
 \begin{equation}
 \frac{V_c}{S_L^0} = \mathcal{V}_c = - \frac{r-1}{1-m} (1-G)    \label{Vc}
 \end{equation}
 For zero gravity, $G=0$, and the typical values $r=6$ and $m=0.3$ for flames, we have stability when $\mathcal{V} < \mathcal{V}_c = - 7.14$. Therefore,   premixed flames with $\mathcal M>0$ can be stabilised if $\mathcal V$ is sufficiently negative in this case. 

It is also worth noting from~\eqref{nondimdisp} that the growth rates of the DL and  RT instabilities, as well as that of the ST instability obviously,    depend on the viscosity ratio $m$.  When $m=1$, corresponding to  flame models where the viscosity is assumed constant, $\mu_2=\mu_1$,  the ST instability represented by the second term in \eqref{nondimdisp} is of course absent. It is also absent when $\mathcal{V}=0$, that is for freely propagating flames. This is so even when viscosity variations are taken into account, i.e. $m \neq 1$, with their effects being still felt through the terms representing the DL and RT instabilities. Furthermore, it is worth emphasising that the analysis of Joulin and Sivashinsky~\cite{joulin1994influence} was carried out for the specific case $\mathcal V=1$, where the dimensional imposed (mean) flow $V$ is equal to $S_L^0$. Their results are often used in the literature for freely propagating flames for which $\mathcal V=0$. The difference in the growth rates between these two cases, is highlighted in the following  formulas:
\begin{align}
    &\mathcal {V}=0:  \qquad \,\, s = \frac{(r-1)(1-G)}{1+m} k - \frac{r+m}{1+m}\mathcal M k^2 \label{V0disp}\\ &\mathcal {V}=1: \qquad  \,\, s = \frac{r-m-(r-1)G}{1+m} k - \frac{r+m}{1+m}\mathcal M k^2, \label{V1disp}
\end{align}
which coincide when the assumption of constant viscosity,  $m=1$, is made. 

\subsection{Improved model for the local propagation speed} \label{sec:ImprovedMark}

In this subsection, we revisit the linear stability analysis,
by highlighting the differences brought about by adopting
propagation speed models other than the simple phenomenological Markstein model~\eqref{SL}.  Such models include those considered in flame studies accounting for the dependence of the flame propagation speed, not only on its curvature, $\nabla \cdot \vect n$,  as first introduced by Markstein~\cite{markstein1988experimental}, but also on the stretching,  $-\vect n\vect n:\nabla \vect v$, induced by the local flow field~\cite[p.132]{batchelor2000introduction}, as advocated by Karlovitz \textit{et. al.}~\cite{karlovitz1953studies}, Eckhaus~\cite{eckhaus1961theory} and Markstein~\cite{markstein1964theory}. Therefore, in general, one may anticipate
an interface propagation model  of the form
\begin{align}
    \frac{S_L}{S_L^0} &=  1+\mathcal{L}_c\, \nabla\cdot \vect n + \left. \mathcal L_s\vect n\vect n:\frac{\nabla\vect v}{S_L^0}\right|_{y=f^-}, \label{SL1}    
\end{align}
involving two Markstein lengths, namely,  $\mathcal{L}_c$ associated with the interface curvature and equal to $\mathcal{L}$ in formula~\eqref{SL}, and $\mathcal{L}_s$ associated with the flow strain.     It is worth mentioning, however,  that the effect of curvature and flow stretching have been shown, for flames modelled by one-step chemistry~\cite{clavin1982effects,matalon1982flames}, to combine into a single quantity known as the flame stretch,  defined as  the  
fractional rate of change of a flame area element~\cite{williams1975review,buckmaster1983lectures,clavin2016combustion} and equal to $-S_L^0 \nabla \cdot \nn - \nn \nn:\nabla \vv$. Under such circumstances, formula~\eqref{SL1} is to be applied with $\mathcal{L}_c=\mathcal{L}_s=\mathcal{L}$.  

It is important to point out that, although   model~\eqref{SL1} is more general than model~\eqref{SL} and that it has been derived rigorously using multi-scale analysis at least for premixed flames~\cite{clavin2011curved},  its applicability to our problem is questionable. This is because it has been derived for flows obeying the Navier--Stokes equation rather than Darcy's equation. It is imperative therefore to examine the applicability of model~\eqref{SL1} in the context of a Darcy's flow. To this end, we have carried out a dedicated multiple-scale analysis for premixed flames with one step chemistry, following the approach of~\cite{clavin1982effects,pelce1988influence,matalon1982flames,clavin1983premixed,keller1994transient}. This rather lengthy analysis, which \blue{is} presented elsewhere~\cite{rajamanickam2024hydrodynamic}, demonstrates that formula~\eqref{SL1} is valid, but in marked contrast to the classical analyses of ~\cite{clavin1982effects,matalon1982flames}, $\mathcal{L}_c \neq \mathcal{L}_s$ in general. Furthermore, $\mathcal L_c$ in Darcy's model, may also depend on $\mathcal V$ and $G$. In view of this finding, it is useful to revisit the linear stability analysis using~\eqref{SL1} instead of~\eqref{SL}. The analysis proceeds exactly as above, except for the following modifications. Specifically, the linearised  interfacial conditions~\eqref{interface00} at $y=0$ now read 
\begin{equation} 
\left \llbracket \frac{\kappa}{\mu}  p'_y  \right \rrbracket  =  \left(\frac{\rho_1}{\rho_2}-1\right) \left(S_L^0 \mathcal{L}_c f'_{xx} \left.+\frac{\kappa_1}{\mu_1}\mathcal L_s p'_{yy} \right|_{y=0^-}  \right)\,, \quad   \llbracket  p'\rrbracket =  \alpha f'  \,, \quad  f'_t = \left. \frac{\kappa_1}{\mu_1}  (\mathcal L_s p'_{yy}  -p'_y)\right|_{y=0^-} + S_L^0 \mathcal{L}_c f'_{xx}  \,,
\label{interface11}
\end{equation}
\blue{which leads to the dispersion relation}
\begin{equation}
    s = \frac{\alpha   (k+\mathcal L_s k^2) - [(\mu_2/\kappa_2)(\rho_1/\rho_2) +\mu_1/\kappa_1]S_L^0\mathcal L_c   k^2}{\mu_1/\kappa_1+\mu_2/\kappa_2 + \mathcal L_s (r-1)k \mu_2/\kappa_2}  \,, \label{dimdisp1}
\end{equation}
 replaceing~\eqref{dimdisp}. Introducing the Markstein numbers $\mathcal M_c =  \mathcal L_c/\delta_L^0$ and $\mathcal M_s = \mathcal L_s/\delta_L^0$, 
the dispersion relation takes the  non-dimensional form
\begin{equation}
   s = \frac{ ak -  b k^2}{1+ ck},  \quad \text {where} \quad   a= \frac{r-1}{1+m} + \frac{1-m}{1+m} \mathcal V + \frac{1-r}{1+m} G, \quad    b = \frac{r+m}{1+m}\mathcal M_c + a \mathcal M_s, \quad    c= \frac{r-1}{1+m}\mathcal M_s \,, \label{nondimdisp1}
\end{equation}
which reduces to~\eqref{nondimdisp} when $\mathcal M_s=0$ and $\mathcal M_c= \mathcal M$.

 \begin{figure}[h!]
\centering
\includegraphics[scale=0.45]{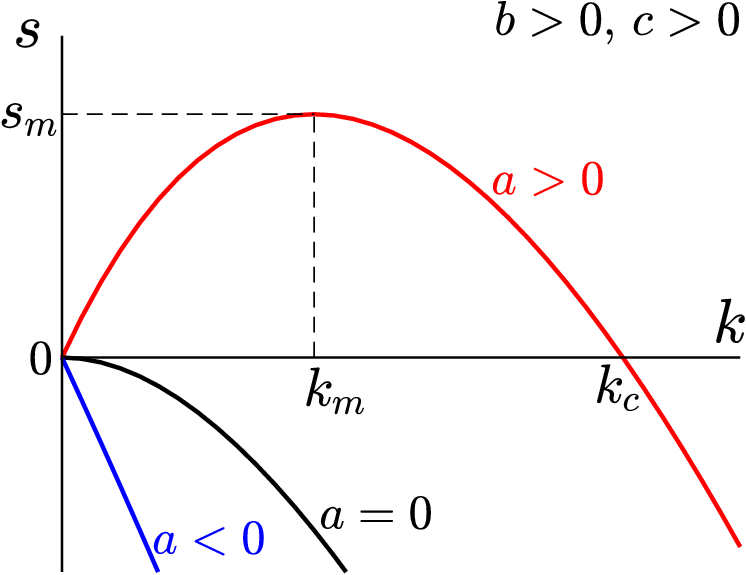}
\caption{Schematic illustration of the formula $s=(ak-bk^2)/(1+ck)$ for $a>0$, $a=0$ and $a<0$, all with $b>0$ and $c>0$. Here, $k_c=a/b$, $k_m = (\sqrt{1+ac/b} -1)/c$ and $s_m=b k_m^2$.} \label{fig:sk1}
\end{figure}

A brief discussion of the main implications of the new dispersion relation~\eqref{nondimdisp1} is now provided. We first note the asymptotic behaviours
\begin{equation}
  s = ak -  k^2(b+ ac) + \cdots \quad \text{as} \quad     k\to 0  \qquad \text{and} 
    \qquad s = -\frac{b}{c}k + \frac{1}{c^2}(b+ac)+\cdots \quad \text{as} \quad k\to \infty \,,
\end{equation} 
which indicates that the parabolic behaviour of the original dispersion relation~\eqref{nondimdisp} is retained for small values of $k$ while a linear behaviour is approached for larger values. We also note that the parameters $b$ and $c$ have to  be both non-negative, which we shall assume, since otherwise $s(k)$ would be unbounded from above for $k \in [0,\infty)$, leading to an ill-posed problem. A schematic illustration of  the dispersion curve $s(k)$ is provided in Fig.~\ref{fig:sk1} for positive values of $b$ and $c$.  Since $b>0$ and $c>0$, the necessary and sufficient condition for instability corresponds to $a>0$ as found earlier, with the instability being hydrodynamic in nature. The maximum of the dispersion curve for the unstable cases is reached at $(k,s)=(k_m,s_m)$ where
\begin{equation}
    k_m = \frac{1}{c}\left(\sqrt{1+\frac{ac}{b}}-1\right), \qquad s_m= bk_m^2.
\end{equation}

 \section{Revisiting Joulin--Sivashinsky analysis based on Euler-Darcy model} \label{modelEulerDarcy}

\subsection{Analysis and discussion} \label{sec:modelEulerDarcyDiscussion}

The pioneering analysis by Joulin and Sivashinsky~\cite{joulin1994influence} was based on an Euler--Darcy model addressing the stability of a premixed flame. This has been  extended recently by Miroshnichenko \text{et. al.}~\cite{miroshnichenko2020hydrodynamic} who included curvature effects characterised by a Markstein number $\mathcal{M}$ while accounting for the presence of an  imposed flow, characterised in our notation by the non-dimensional parameter $\mathcal{V}$. In the  analysis of~\cite{miroshnichenko2020hydrodynamic}  gravity effects were neglected and the unburnt to burnt density ratio ($r$ in  our notation)  and viscosity ratio ($m$) were lumped together into a single parameter, which somewhat obscures their relative contributions. We shall briefly revisit the  Euler--Darcy model relaxing the assumptions made in~\cite{joulin1994influence, miroshnichenko2020hydrodynamic}, to match those of the Darcy's model of the previous section.  In this subsection, we shall adopt for simplicity the Markstein model~\eqref{SL}, and provide  in this context   a short derivation of the dispersion relation  along with a discussion of its implications. 
For completeness,  the dispersion relation  corresponding to the improved propagation speed model~\eqref{SL1} will be provided in the following subsection. 

The analysis follows closely that of the previous section. The governing equations in the laboratory frame, written on both sides of the interface, are  now given by   
\begin{equation}
    \nabla\cdot\vect u = 0 \,, \qquad \rho \frac{d\vect u}{dt} = -  \nabla (p+\rho gy) - \frac{\mu}{\kappa}\vect u \,, \label{eulerdarcylab}
\end{equation}
instead of Darcy's model~\eqref{darcylab}. In the moving frame introduced above and illustrated in Fig.~\ref{fig:config}(b), the equations, in terms of  $\vect v=\vect u + (S_L^0-V)\vect e_y$, take the form 
\begin{equation}
    \nabla\cdot\vect v = 0 \,, \qquad \rho \frac{d\vect v}{dt} = -  \nabla (p+\rho gy) - \frac{\mu}{\kappa}[\vect v-(S_L^0-V)\vect e_y] \,. \label{eulerdarcymove}
\end{equation}
These are  subject to the boundary condition $\vv = S_L^0 \,  \vect e_y$ as $ y \to - \infty$, and the following familiar interfacial conditions~\cite{matalon2018darrieus} at  $y=f(x,t)$: 
\begin{equation}
    \llbracket \vect v\cdot\vect n \rrbracket = \left(\frac{\rho_1}{\rho_2}-1\right) S_L, \qquad \llbracket p\rrbracket = -\left(\frac{\rho_1}{\rho_2}-1\right)\rho_1 S_L^2, \qquad \llbracket\vect v\times\vect n\rrbracket =0, \qquad S_L = \left. \frac{v-u f_x -f_t}{\sqrt{ 1+f_x^2}} \right |_{y=f^-}. \label{interfaceeulerdarcy}
\end{equation}
These conditions replace those in~\eqref{interfacedarcy} and differ by the presence of the third condition which expresses the continuity of the tangential component of $\vv$, a requirement which is not needed or enforceable for the Darcy's law. Furthermore, there is a jump in pressure  across the interface characterised by the second condition in~\eqref{interfaceeulerdarcy} associated with the presence of the inertial term on the left hand side of~\eqref{eulerdarcymove}; this jump is absent in~\eqref{interfacedarcy}.

The basic solution   $(\vv,p,f)=\left(\overline \vv,  \overline{p},\overline{f}\right)$
corresponding to a stationary flat interface  is given by $\overline f =0$ and
\begin{align} 
 &\overline \vv=S_L^0\vect e_y\,, \quad \quad  -\overline p= \rho_1 gy   + \frac{\mu_1}{\kappa_1}Vy \quad   \text{for} \quad  y<0  \,, \label{basicEq1} \\ 
& \overline \vv =\frac{\rho_1}{\rho_2}S_L^0\vect e_y \,,  \quad  -\overline p=  \rho_2 gy + \frac{\mu_2}{\kappa_2}Vy +  \left(\frac{\rho_1}{\rho_2}-1\right)\left(\frac{\mu_2}{\kappa_2}Vy + \rho_1 S_L^{0}\right) S_L^{0}\quad \text{for} \quad  y>0  \,. \label{basicEq2}
\end{align}
As shown in Appendix~\ref{appendixa}, the stability of this solution is found to governed by the quadratic dispersion relation
\begin{equation} \label{dimdispeulerdarcy}
     (\rho_1+\rho_2) s^2 +\left[\frac{\mu_1}{\kappa_1}+ \frac{\mu_2}{\kappa_2}  + 2\rho_1S_L^0(1+\mathcal L k) k\right]s - \alpha  k + \left[  \left( \frac{\rho_1}{\rho_2 } \frac{\mu_2}{\kappa_2}+ \frac{\mu_1}{\kappa_1} \right) \mathcal{L}-    \left( \frac{\rho_1}{\rho_2}-1\right)\rho_1 {S_L^0}\right]S_L^0 k^2 +    \frac{2 \rho_1^2 {S_L^0}^2 \mathcal{L}}{\rho_2}  k^3=0 \,,
\end{equation}
where $\alpha$ is as given in~\eqref{alpha}. A non-dimensional form of this dispersion relation  is obtained by multiplying all terms by ${\delta_L^0}^2/ \rho_1  {S_L^0}^2$ and introducing the non-dimensional growth rate and wave number  $\tilde{s}= s \delta_L^0/S_L^0$ and $\tilde{k}= k \delta_L^0$ and using the parameters defined in~\eqref{nondim}.   Dropping the tildes we thus obtain the non-dimensional dispersion relation
\begin{equation}
  \frac{r+1}{r} s^2 +           \left[2    \left(1+ \mathcal{M} k\right)k  + \varphi   \right] s                     - a \varphi   k + \left(   b \varphi+  1- r    \right)  k^ 2+     2 r  \mathcal{M} k^ 3=0 \,, \label{nondimdispeulerdarcy}
\end{equation}
where  $a$ and $b$ are as defined in~\eqref{nondimdisp} and
\begin{equation} \label{varphi}
\varphi = \frac{12 \Pra(1+m)}{\ep^2m} \qquad \text {with} \quad \ep = \frac{\sqrt{12 \kappa_1}}{\delta_L^0} \quad \text{and} \quad \Pra= \frac{\mu_1}{\rho_1 D_1} \,. 
\end{equation}
 We note that for a Hele-Shaw channel,  $\kappa_1 = h^2/12$ so that $\epsilon=h/\delta_L^0$.  In the narrow channel limit $\epsilon \to 0$, terms containing $\varphi$ are dominant so that dropping all other  terms implies $s=ak-bk^2$ which is the non-dimensional dispersion relation~\eqref{nondimdisp} obtained above using Darcy's law. In the wide-channel limit, $\ep\to \infty$,
 $\varphi \to 0$, leading to a classical result first obtained by Markstein~\cite{markstein1988experimental} for flames freely propagating in unconfined geometries. In the latter context, more   accurate dispersion relations are available, accounting for temperature dependence of all transport coefficients~\cite{clavin1983influence, matalon2018darrieus}, but these are not directly applicable in narrow confined channels which are our main focus. Furthermore, in wide channels, the dependence of (the effective speed) $S_L$ in~\eqref{SL} on $V$ cannot strictly speaking  be ignored~\cite{daou2001flame,miroshnichenko2020hydrodynamic}  as we assumed herein~\footnote{In particular, for premixed flames in wide channels we may replace~\eqref{SL}, motivated by the study in \cite{daou2001flame}, by the approximate formula $S_L/S_L^0=(1+ \mathcal V)(1 + \mathcal L\nabla\cdot \nn) $ for $\mathcal V>0$ and $S_L/S_L^0=(1-\mathcal V/2 )(1 + \mathcal L\nabla\cdot \nn)$ for $\mathcal V<0$. For small channels, relevant to our study, as shown in~\cite{daou2018taylor,rajamanickam2023thick,rajamanickam2024effect} the appropriate formula is given by $S_L/S_L^0 = (1+\mathcal L\nabla\cdot\vect n)(1+\gamma \ep^2 m^2 \mathcal V^2/r^2)/\sqrt{1+\gamma \ep^2 m^2 \mathcal V^2\Lew^2/r^2}$, where $\Lew$ is the Lewis number and $\gamma$ is the Taylor-dispersion coefficient, which takes the value $\gamma=1/210$ for a Poiseuille flow. The corrections introduced here
 are equivalent to simply redefining $S_L^0$ in~\eqref{SL} and elsewhere by multiplying by a constant.  With the redefined $S_L^0$, the dispersion relation~\eqref{nondimdispeulerdarcy} is still applicable, without the need for a separate stability analysis.}.
 
To close this section, we derive now a simple stability criterion based on \eqref{nondimdispeulerdarcy}, assuming as before that the Markstein number $\mathcal M>0$.  Then, the coefficients of $s^2$ and $s$  in~\eqref{nondimdispeulerdarcy} are clearly positive. Therefore, by Routh--Hurwitz criterion for  the second-order polynomial in $s$,  instability occurs if and only if  the constant coefficient of the quadratic equation is negative. This condition is equivalent to finding whether the minimum value of the parabola 
\begin{equation}
    g( k) =  2r\mathcal Mk^2 + (b\varphi+1-r)k - a \varphi 
\end{equation}
in the interval $ k\in [0,\infty)$  is negative. The minimum occurs at $ k= k_* \equiv (r-1-b\varphi)/4r\mathcal M$ and its value is $g=g_* \equiv - a \varphi  -2r\mathcal M k_*^2$. For stability, we then require $g_*>0$, i.e., $a \varphi<-2r\mathcal M  k_*^2$, and this is applicable as long as $ k_*>0$. When $ k_*<0$, the condition for stability is simply given by $g(0)>0$, which requires $a<0$. This second stability criterion  coincides with the stability criterion  for Darcy's law, whereas the first criterion pertains to cases with $\varphi$ (or $1/\ep$) being sufficiently small leading to  $ k_*$ being positive when $r>1$.  This suggests that the predictions of the Darcy and Darcy-Euler models agree with each other provided $k_*<0$, a condition which may be written  
as $\epsilon < \epsilon_c$  where
\begin{equation} \label{epc}
\epsilon_c^2 = \frac{12 Pr \,  \mathcal{M} (r+m) }{m(r-1)}   \,.
\end{equation}
Note that both  criteria may be combined to yield the necessary and sufficient condition for  stability in the form 
\begin{equation} \label{ac}
     a< a_c \equiv 
     \begin{cases}\begin{aligned}
        &0   \qquad \qquad \qquad \qquad \qquad \qquad \text{for} \quad \ep^2 < \epsilon_c^2 \,,\\
       &-\frac{b(r-1)}{8 r\mathcal{M}} \, \frac{\ep^2}{\ep_c^2}\left(1-\frac{\ep_c^2}{\ep^2}\right)^2  \, \, \quad \text{for} \quad \ep^2 > \epsilon_c^2 \,.
    \end{aligned}\end{cases}
\end{equation}
Note that the stability criterion is always $a<a_c=0$ when
$r<1$, since then $k_m$ is clearly negative. When $r>1$ as  for premixed flames,  both cases in the stability criterion~\eqref{ac} occur. For premixed flames, we can use the typical values $\Pra=0.7$, $r=6$ and $m=0.3$ to evaluate the coefficients in~\eqref{ac} which implies that   $\ep_c = 6\sqrt{\mathcal M}$ and $b(r-1)/8r\mathcal M = 0.5$. The stability criteria can also be written in terms of the critical velocity $\mathcal V_c$ below which the interface is stable. This critical value  is given by~\eqref{Vc} for Darcy's law, and this remains true for the Euler--Darcy model  when $\ep<\ep_c$, which  always holds for $r<1$. When  $\ep>\ep_c$,  $\mathcal V_c$ is to be computed using the second case in formula~\eqref{ac}. For $r>1$, we thus find 
\begin{equation}
    \mathcal V_c =  \begin{cases}\begin{aligned}
        & - \frac{r-1}{1-m}(1-G)   \qquad \qquad \qquad \qquad \qquad \qquad \,\, \text{for} \quad \ep  < \epsilon_c  \,,\\
       &- \frac{r-1}{1-m}\left[1-G+\frac{r+m}{8 r} \, \frac{\ep^2}{\ep_c^2}\left(1-\frac{\ep_c^2}{\ep^2}\right)^2 \right] \, \, \quad \text{for} \quad \ep > \epsilon_c  \,.
    \end{aligned}\end{cases}
\end{equation}
We note that $\mathcal V_c \to -\infty$ as $\ep \to \infty$, indicating that a flame in an infinitely wide channel is impossible to stabilise by the flow.

\subsection{The dispersion relation  for Darcy--Euler with an improved interface-propagation model}

The previous subsection was based on the dispersion relation~\eqref{dimdispeulerdarcy} and its non-dimensional form~\eqref{nondimdispeulerdarcy}, obtained when adopting the Markstein's propagation speed  model~\eqref{SL}. In this subsection,  we provide for completeness the dispersion relation corresponding to the  improved propagation speed model~\eqref{SL1}, whose derivation is briefly outlined at the  end of Appendix~\ref{appendixa}. In dimensional form, the dispersion relation is found to be given by 
\begin{equation}  
       a_2 s^2 +a_1 s + a_0=0   
\end{equation} 
where
\begin{eqnarray*}
  a_2 &=& \rho_1+\rho_2 + (\rho_1-\rho_2) \mathcal L_sk \,,\\
  a_1 &=& \frac{\mu_1}{\kappa_1}+ \frac{\mu_2}{\kappa_2}  + \left[2\rho_1S_L^0 + \frac{\mu_2}{\kappa_2} \left(\frac{\rho_1}{\rho_2}-1\right)\mathcal L_s\right]k  + 2\rho_1S_L^0 \left[ \mathcal L_c+  \left(\frac{\rho_1}{\rho_2}-1\right)\mathcal L_s\right] k^2 \,,\\
  a_0 &=& - \alpha  k + \left[\left(\frac{\rho_1}{\rho_2 } \frac{\mu_2}{\kappa_2}+ \frac{\mu_1}{\kappa_1} \right) S_L^0\mathcal L_c +    \left(1- \frac{\rho_1}{\rho_2}\right)\rho_1 {S_L^0}^2 + \alpha \mathcal L_s\right] k^2 +    \frac{2 \rho_1^2 {S_L^0}^2 }{\rho_2} \left[\mathcal L_c + \left(1-\frac{\rho_2}{\rho_1}\right) \frac{\mathcal L_s}{2}\right] k^3.
\end{eqnarray*}
 As done above, a non-dimensional form is obtained by multiplying all terms by ${\delta_L^0}^2/ \rho_1  {S_L^0}^2$ and introducing the non-dimensional growth rate and wave number  $\tilde{s}= s \delta_L^0/S_L^0$ and $\tilde{k}= k \delta_L^0$. Dropping the tildes, we thus have  
\begin{eqnarray} \label{eq:DarcyEulerDispImproved}
  \frac{r+1 +(r-1) \mathcal M_s k }{r} s^2 &+&          \big[2 k   \big\{1+ \left[\mathcal M_c +(r-1)\mathcal{M}_s \right] k\big\}  + \varphi\left( 1+c  k \right)   \big] s                    \\
  &-& a \varphi   k + \big(   b \varphi+  1- r    \big)  k^ 2+   \big[  2 r   \mathcal{M}_c + (r-1) \mathcal{M}_s \big] k^ 3 = 0 \nonumber
\end{eqnarray}
where  $a$, $b$ and $c$ are as defined in \eqref{nondimdisp1} and $\varphi$, $\epsilon$, $Pr$ as defined in \eqref{varphi}. It is worth noting that in the narrow channel limit  $\epsilon \to 0$, terms containing $\varphi$ are dominant so that dropping all other  terms  leads to the dispersion relation~\eqref{nondimdisp1} derived for Darcy's law. On the other hand, as $\epsilon \to \infty$, the terms \blue{containing $\varphi$} drop, and we obtain exactly the same dispersion relation as the one derived as equation~(10) in the paper by Creta and Matalon~\cite{creta2011strain} provided $\mathcal M_s = \mathcal M_c$. 

 \begin{figure}[h!]
\centering
\includegraphics[width=0.8\textwidth]{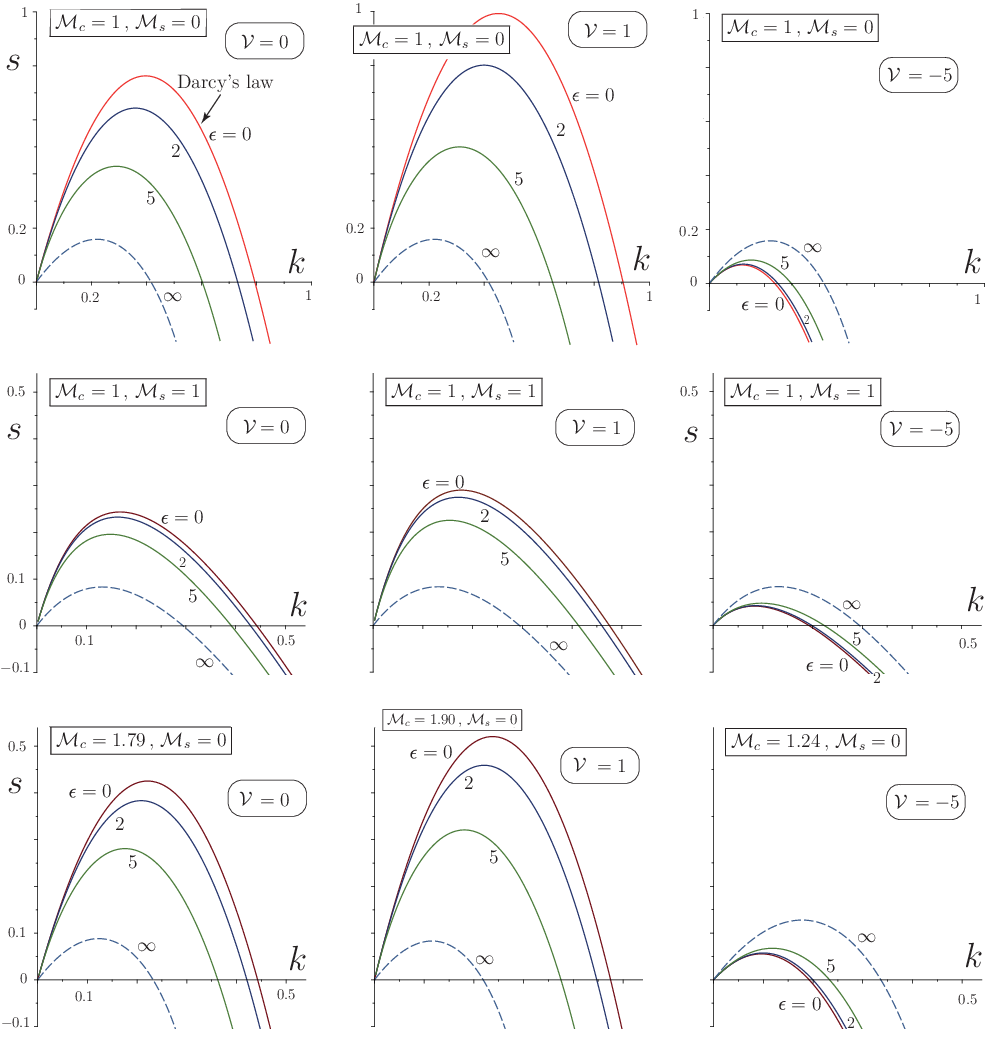}
\caption{The growth rate $s$ versus the wavenumber $k$ based on the dispersion relation~\eqref{eq:DarcyEulerDispImproved}. All calculations are performed with the parameter values  $r=6$, $m=0.3$, $G=0$ and $Pr=0.7$.  The three cases in the top row have Markstein numbers $\mathcal M_c=\mathcal M_s=1$.  The cases in the bottom row  have all $\mathcal M_s=0$ and $\mathcal M_c$ as indicated in each subfigure.  The dispersion curves correspond to three values of $\mathcal{V}$, namely $0$, $1$ and $-5$ and selected values of $\epsilon$, as indicated. \blue{The axes of the subfigures in each row have the same horizontal and vertical length scales and labels.}} \label{fig:SvskAllFigures}
\end{figure}

Illustrative results based on the dispersion relation~\eqref{eq:DarcyEulerDispImproved} are shown
in Fig.~\eqref{fig:SvskAllFigures}, for  the parameter values  $r=6$, $m=0.3$, $Pr=0.7$ and $G=0$.
The three subfigures in the top row are characterised by the Markstein numbers $\mathcal M_c= 1$ and $\mathcal M_s=0$, that is correspond to cases using the Markstein model~\eqref{SL} where the effect of strain  on the interface propagation speed is ignored.  In each subfigure, the value of $\mathcal V$ is prescribed for selected values of $\epsilon$. When $\epsilon \to 0$, the  curve corresponding to the dispersion relation~\eqref{nondimdisp} based on  Darcy's law and dependent on $\mathcal V$  is approached, while as $\epsilon \to \infty$, the curves  tend to that of the classical Darrieus--Landau instability which is independent of $\mathcal V$. The effect of confinement is prominent for moderate and smaller values of $\epsilon$,   being  destabilising  for larger positive values of $\mathcal V$.   

Turning now to the middle row of Fig.~\eqref{fig:SvskAllFigures},   the effects of curvature and strain  are both retained, corresponding to the improved propagation speed model~\eqref{SL1}  with $\mathcal M_c =\mathcal M_s=1$. Comparing with  the top row, two observations are in order.  First, the range of unstable modes  and the maximum growth rates are now significantly reduced. Second, the parabolic shape of the dispersion curves of the first row is now strongly modified for larger values of $k$, in line with the discussion at the end of subsection~\ref{sec:ImprovedMark}. This change in shape is intimately associated with non-zero values of $\mathcal M_s$ and cannot be avoided by adjusting the curvature Markstein number $\mathcal M_c$ while using the Markstein model~\eqref{SL}. This is evident in the dispersion curves presented in the bottom row. These curves were calculated for $\mathcal M_s=0$ , with $\mathcal M_c$ adjusted to maintain the same range of unstable modes observed in the middle row subfigures.

\section{Darrieus--Landau and Saffman--Taylor instabilities based on Darcy's law: Numerical results for a premixed flame}\label{numerical}

\subsection{Problem formulation}
In the  theoretical analysis of the previous sections,  the interface is treated as a hydrodynamic discontinuity in the spirit of the original studies of the Darrieus--Landau and Saffman--Taylor instabilities~\cite{darrieus1938propagation,landau1944slow,saffman1958penetration,rayleigh1882investigation,taylor1950instability}. The analysis also accounted  for local curvature effects of the interface, following the phenomenological approach of Markstein~\cite{markstein1988experimental}. Strictly speaking, a more elaborate description  of curvature effects is possible, accounting for the internal  structure of the interface, as found for premixed flames in~\cite{matalon1982flames,clavin1983premixed}. In the remainder of the paper, we shall present a numerical investigation focusing on the hydrodynamic instabilities of premixed flames in order to complement and validate the theoretical results of the Darcy's law model. 

We shall adopt a 2D configuration for our investigation as illustrated in Fig.~\ref{fig:config},  which may be associated with a depth-averaged description of flames in a Hele-Shaw channel. In this configuration, fluid 1 corresponds to an unburnt reacting mixture and fluid 2 to a burnt gas mixture. For simplicity,  the unburnt gas is assumed  to be deficient in a reactant which is consumed with a reaction rate  per unit volume  given by an  Arrhenius law $\rho B Y_R e^{-E/RT}$. Here, $B$ is the pre-exponential factor, $Y_R$ the reactant mass fraction,  $T$ the  gas temperature, $E$ the activation energy of the reaction, and $R$ the universal gas constant.   For this model, we may define the adiabatic flame temperature by $T_{2}=T_1(1+q)$, where $q$ quantifies the amount of heat released by the reaction and $T_1$ the unburnt gas temperature. A non-dimensional measure of the activation energy is then given by  the Zeldovich number $\beta = E(T_{2}-T_1)/RT_{2}^2$.  Further, we shall assume  that the thermal diffusivity $D$ is equal to the reactant diffusion coefficient, i.e.~a unit Lewis number, so as to eliminate \blue{diffusive-thermal} (Turing-like) instabilities. We shall also ignore heat losses, whether these are by radiation,  or by conduction to walls. These assumptions are adopted for simplicity in order to focus on the flame hydrodynamic instabilities.  Under these conditions,   the  temperature $T$ and the   mass fraction $Y_R$ are not independent, but are related by the equation $Y_R/Y_{R,1} =1-(T/T_1-1)/q$, where $Y_{R,1}$ is the mass fraction in the unburnt gas. Finally, the density  is assumed to depend on temperature according to the ideal gas law, while the transport coefficients are assumed to follow a power-law temperature dependence. Specifically, 
\[
\rho T = \rho_1 T_1 \,, \qquad \frac{\mu}{\mu_1} = \frac{\rho D}{\rho_1 D_1} = \left(\frac{T}{T_1}\right)^n  \quad \text{with} \quad n=0.7 \,.
\]

For convenience, we introduce the following non-dimensional variables and parameters:
\begin{align}
    t^* = \frac{tS_L^0}{\delta_L}, \quad \vect x^* = \left(\frac{x}{\delta_L},\frac{y}{\delta_L}\right), \quad \rho^* = \frac{\rho}{\rho_1}, \quad \vect u^*= \frac{\vect u}{S_{L}^0}, \quad p^* = \frac{h^2p}{12\mu_1D_{1}}, \quad \theta = \frac{T-T_1}{T_2-T_1}, \nonumber \\  \mu^* = \frac{\mu}{\mu_1}, \quad \lambda = \frac{\rho D}{\rho_1 D_{1}},  \quad \mathcal V = \frac{V}{S_L^0},\quad G = \frac{\rho_2 gh^2}{12\mu_2 S_L^0},  \quad r = \frac{\rho_1}{\rho_2}, \quad m = \frac{\mu_1}{\mu_2}, \quad S = \frac{S_L^0}{S_{L,\infty}^0}, \nonumber
\end{align}
where $\delta_L=D_1/S_L^0$ is the laminar flame thickness and $S_{L,\infty}^0=  [2\beta^{-2} B  D_1  (\rho_2^2 D_2)/(\rho_1^2 D_1)e^{-E/RT_2}]^{1/2}$ is the asymptotic formula for $S_L^0$ derived in the limit $\beta\to\infty$~\cite[p.164]{williams2018combustion}. The value of $S$, which must approach unity as $\beta\to \infty$, will be determined (by solving~\eqref{basetheta} below) numerically for the typical finite value $\beta = 10$, adopted herein.  Dropping the asterisks for $t^*$, $\vect x^*$, $\rho^*$, $\vect u^*$, $p^*$ and $\mu^*$, the two-dimensional governing equations in the laboratory frame, can be written as
\begin{align}
    \pfr{\rho}{t} + \nabla\cdot(\rho \vect u) & = 0, \label{labcont}\\
    -\mu \vect u  = \nabla p &+  \frac{rG}{m}\rho\,\vect e_y , \\
    \rho \pfr{\theta}{t} + \rho \vect u \cdot \nabla \theta  & = \nabla\cdot(\lambda \nabla\theta) + \omega, \label{thetaflamelab}\\    
     \rho(1+q\theta)=1, &\quad \mu = \lambda= (1+q\theta)^n  \label{eqn}
\end{align}
where
\begin{equation}
    \omega = \frac{\beta^2}{2S^2} (1+q)^{1-n} \rho (1-\theta) \exp\left[\frac{\beta(\theta-1)}{ 1+q (\theta-1)/(1+q)}\right]. \label{omega}
\end{equation}
From equation~\eqref{eqn}, it follows that the density ratio $r=1+q$ and the viscosity ratio $m=1/(1+q)^{n}$, since $\theta \to 1$ in the burnt gas as $y\to +\infty$ and $\theta \to 0$ in the unburnt gas as $y\to -\infty$.

\subsection{Linear stability of the planar flame} \label{sec:linearnumerical}

In the laboratory frame, the planar flame propagates in the negative $y$-direction with a constant  non-dimensional speed $1-\mathcal V$ since $\vect u = \mathcal V \vect e_y$ as $y\to-\infty$; see Fig.~\ref{fig:config}. To study the stability of this planar flame, it is advantageous to shift to a coordinate system moving with the  flame front, by using the coordinate transformation   $(x,y,t)\mapsto (x,y+(1-\mathcal V)t,t)$. Introducing further  $\vect v = \vect u + (1-\mathcal V)\vect e_y$, we have $\vect v= \vect e_y$ as $y\to-\infty$ and the governing equations may be written as
\begin{align}
  -\frac{1}{q}\nabla\cdot \left(\frac{\nabla p}{\mu}\right) - \frac{rG}{mq}\pfr{}{y} \left(\frac{\rho}{\mu}\right) &= \nabla\cdot(\lambda \nabla\theta) +\omega, \label{Poisson}\\
    -\mu \vect v + \mu (1-\mathcal V) \vect e_y  &= \nabla p +\frac{rG}{m}\rho\,\vect e_y,  \label{darcy1}\\
    \rho \pfr{\theta}{t} + \rho \vect v \cdot \nabla \theta   & = \nabla\cdot(\lambda \nabla\theta) + \omega, \label{theta1}\\    
     \rho(1+q\theta)=1, &\quad \mu = \lambda= (1+q\theta)^n. \label{eqn1}
\end{align}
Instead of the continuity equation, we have introduced the Poisson equation~\eqref{Poisson} for the pressure field, which is obtained by  combining the continuity equation $\pfi{\rho}{t}+ \nabla\cdot(\rho\vect v)=0$, the temperature equation~\eqref{theta1} and the equation of state given in~\eqref{eqn1}.

The basic solution, corresponding to the planar flame,  is governed by equations~\eqref{Poisson}-\eqref{eqn1} in which   the time and $x$-derivatives are set to zero. Denoted with an overbar, it is given by 
\begin{align}
    \overline\rho\, \overline{\vect v} = \vect e_y \,, \quad -\frac{d\overline p}{dy} =  \overline\mu  \frac{1-\overline\rho}{\overline\rho} + \overline\mu \mathcal V +\frac{rG}{m} \overline\rho \,, \label{baseflow}
\end{align}
along with $\overline\rho=1/(1+q\overline\theta)$ and $\overline \mu = \overline\lambda= (1+q\overline\theta)^n$, where $\overline\theta$ satisfies
\begin{align}
     \frac{d\overline \theta}{dy} = \frac{d}{dy}\left(\overline\lambda \frac{d\overline\theta}{dy}\right)  + \omega(\overline \theta),   \quad \overline\theta(-\infty)=0, \quad \overline\theta(+\infty)=1 \,. \label{basetheta}   
\end{align}
The temperature field $\overline\theta(y)$  and the unknown parameter $S$ which  appears in the definition of $\omega$ in~\eqref{omega}, are independent of $\mathcal V$ and $G$, and are computed numerically for the typical values $n=0.7$, $q=5$ and $\beta=10$; for these values, $S=0.929$.

The stability of the planar flame  is investigated by introducing infinitesimal perturbations   such that
\begin{equation}
\begin{bmatrix}
  \theta \\  
 p
\end{bmatrix}=
    \begin{bmatrix} \,
\overline \theta(y) \\
\, \overline p(y)
\end{bmatrix}+ \begin{bmatrix}
\hat \theta(y) \\
\hat p(y)
\end{bmatrix}e^{st+ikx} \label{spannormal}
\end{equation}
where $k$ is the real wavenumber of the perturbation and $ s$ its growth rate, which is to be obtained as an eigenvalue. Substituting~\eqref{spannormal} into~\eqref{Poisson}-\eqref{eqn1} and linearising about the base solutions~\eqref{baseflow}-\eqref{basetheta} results in the linearised system of equations 
\begin{align}
      s \overline \rho \hat \theta + \frac{d\hat\theta}{dy} + \overline\rho(\hat v_y -q\hat\theta) \frac{d\overline\theta}{dy} & = \frac{d}{dy}\left(\overline\lambda \frac{d\hat\theta}{dy}+\hat\lambda \frac{d\overline\theta}{dy}\right) -  k^2 \overline\lambda \hat \theta + \hat \omega \,, \label{lineartheta}\\  
  - \frac{1}{q} \frac{d}{dy}\left(\frac{1}{\overline\mu} \frac{d\hat p}{dy}- \frac{\hat\mu}{\overline\mu^2}\frac{d\overline p}{dy}\right) + \frac{ k^2 \hat p}{q \overline\mu}  - \frac{rG}{mq} \frac{d}{dy}\left(\frac{\hat \rho}{\overline\mu} - \frac{\overline\rho \hat\mu}{\overline\mu^2}\right)& = \frac{d}{dy}\left(\overline\lambda \frac{d\hat\theta}{dy}+\hat\lambda \frac{d\overline\theta}{dy}\right) -  k^2 \overline\lambda \hat \theta + \hat \omega \,. \label{linearp}
\end{align}
Here $\hat \rho = -q\overline\rho^2\hat\theta$, $\hat\mu = nq \overline\rho\,\overline\mu \hat \theta$,  $\hat\lambda = nq \overline\rho\,\overline\lambda \hat \theta$, $-\overline \mu\hat v_y=d\hat p/dy + \hat \mu(q\overline\theta+\mathcal V) + \hat \rho rG/m$ and 
\begin{align} 
   \hat\omega=  \frac{\beta^2 (1+q)^{1-n}}{2 S^2} 
    \left[\hat\rho(1-\overline\theta) - \overline\rho\hat \theta + + \frac{\overline\rho (1-\overline\theta) \beta(1+q)^2\hat\theta}{(1+q\overline\theta)^2}\right] \exp\left[\frac{\beta(\overline \theta-1)}{ 1+q (\overline \theta-1)/(1+q)}\right] \,.  
\end{align}
Solving the linear eigenvalue problem given by  equations~\eqref{lineartheta}-\eqref{linearp} subject to the boundary conditions $\hat\theta=d\hat p/dy =0$ as $y\to\pm\infty$ yields numerically the required dispersion relation
\begin{equation}
    s = s (k;\mathcal V,G) \,.   \label{dispnumerical}
\end{equation}

\subsection{The dispersion relation computed numerically and its implications}

\begin{figure}[h!]
\centering
\includegraphics[scale=0.55]{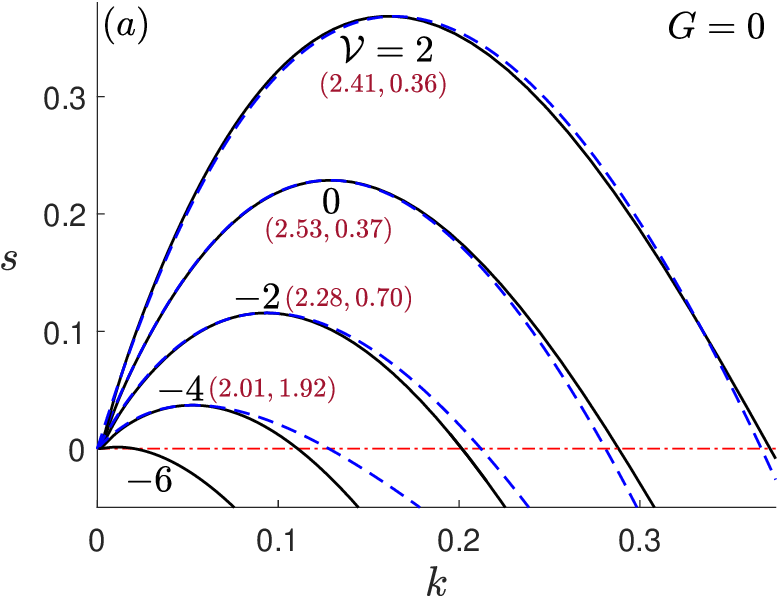}\hspace{1.5cm}
\includegraphics[scale=0.55]{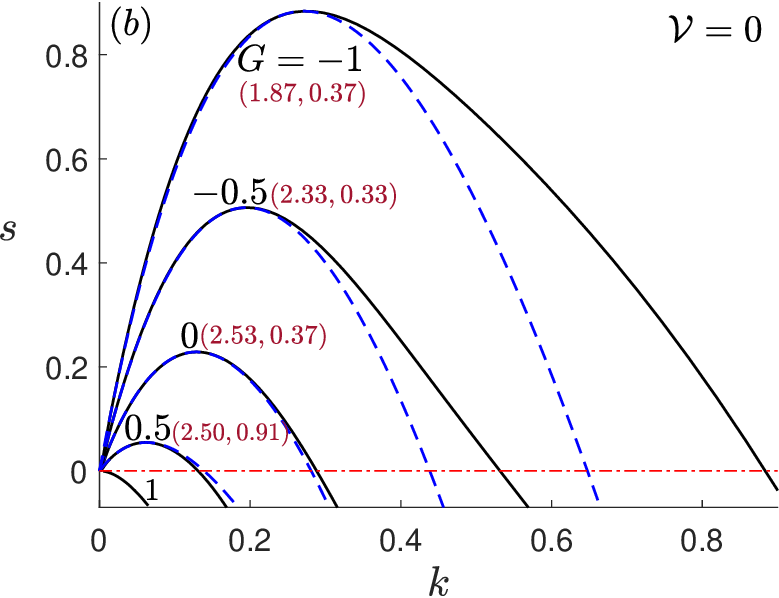}
\caption{Dispersion curves $s(k)$  represented as solid lines computed for   $q=5$, $n=0.7$ and $\beta=10$ (i.e.,~$r=6$ and $m=0.29$). Left figure (a) corresponds to $G=0$ and selected values of $\mathcal V$.
Right figure (b) corresponds to $\mathcal V=0$ and selected values of $G$.  The dashed lines correspond to the theoretical formula $s=(ak -b k^2)/(1+ck)$ with $a$ evaluated from equation~\eqref{nondimdisp1}, while  $b$ and $c$ are evaluated using the relations $b=s_m/k_m^2$ and $c=a/s_m-2/k_m$, where $(k_m,s_m)$ corresponds to the maximum point of the numerical dispersion relation. The numbers inside the parentheses for each curve correspond to the resulting values of $\mathcal M_c$ and $\mathcal M_s$ extracted from $b$ and $c$.} \label{fig:dispVG}
\end{figure}

The dispersion relation~\eqref{dispnumerical} is computed numerically for the typical values  $q=5$, $n=0.7$ and $\beta=10$. The computations are carried out using the eigenvalue solver in COMSOL Multiphysics, as done in our earlier works~\cite{daou2024diffusive,rajamanickam2023stability,kelly2023influence,kelly2024three}. Illustrative numerical results showing the growth rate $s$ versus the wavenumber $k$  for selected values of the parameters  $\mathcal V$  and $G$ are plotted in  Fig.~\ref{fig:dispVG} as solid lines. Fig.~\ref{fig:dispVG}(a), corresponding to cases without gravity ($G=0$), shows that the perturbation growth rate increases with increasing values of $\mathcal V$. As $\mathcal V$ is decreased, the maximum growth rate  decreases and the range of unstable wavenumbers shrinks. For $\mathcal V=-6$, $s(k)<0$ except for values of $k$ in a tiny range (difficult to see in the figure) around $k=0$.  For lower values of  $\mathcal V$,  the growth rate is non-positive for all wavenumbers. This demonstrates the existence of   a critical value of $\mathcal V$  below which the flame is stable, as predicted theoretically in formula~\eqref{Vc} of Section~\ref{modelDarcy}. A similar trend can also be observed in Fig.~\ref{fig:dispVG}(b) where $G$ is varied while maintaining $\mathcal V=0$ which corresponds to freely propagating flames. Here, as expected, an increase in the value of $G$ is stabilising and there exists a critical value of $G$ above which the flame is stable, again in agreement with the theoretical results discussed in Section~\ref{Sec:Implications}.

The dashed lines in Fig.~\ref{fig:dispVG} represent the theoretical formula $s =(ak - bk^2)/(1+ck)$, where $a$ is evaluated using the formula given in~\eqref{nondimdisp1}.  As for the parameters $b$ and $c$ which are also defined in~\eqref{nondimdisp1}, their evaluation requires the values for the two Markstein numbers $\mathcal M_c$ and $\mathcal M_s$,
which are not readily available from existing theories, as the latter  applicability to Darcy's flows has not been established.  The determination of these Markstein numbers, which may also be influenced by $\mathcal V $ and $G$ in the context of Darcy's flows as a preliminary analysis suggests, requires dedicated investigations, say  multiple-scale analysis accounting for the  flame's inner structure or numerical/experimental approaches. For the sake of assessing the ability
of our theory to predict flame instability,  we shall  herein  evaluate these parameters directly from the numerical dispersion relation itself, as sometimes done in experimental studies~\cite{al2019darrieus}. 
To this end, we shall assume $b=s_m/k_m^2$ and $c=a/s_m-2/k_m$ with $(k_m,s_m)$ corresponding to the maximum location of the numerical dispersion curve, so that the growth rates from the theoretical formula $s=(ak-bk^2)/(1+ck)$   and from  the numerical dispersion relation~\eqref{dispnumerical} coincide at the maximum location. It follows that $\mathcal M_s = c(1+m)/(r-1)$ and $\mathcal M_c=(b-a\mathcal M_s)(1+m)/(r+m)$, and these are reported in Fig.~\ref{fig:dispVG} inside the parentheses for each curve.  As can be inferred then from Fig.~\ref{fig:dispVG},  the numerical results corroborate the theoretical formula reasonably well, once the
location of the maximum of the numerical and theoretical growth rates are  fitted. In particular, the shape of the numerical dispersion curve is well reproduced by the theory. Note however that the agreement 
between the solid and dashed curves appears to be less satisfactory for larger values of $k$ in  the two cases of Fig.~\ref{fig:dispVG}(b) corresponding to $G=-1$ and $G=-0.5$.  This deterioration of the agreement is not surprising, given that our modelling of the flame as a hydrodynamic discontinuity is, strictly speaking, valid  when $k\ll 1$, 
that is for  perturbations with  wavelength larger than the flame thickness.   More importantly, the parameter $a$, which
characterises the slope of $s(k)$ at $k=0$,  is very well predicted by the theory and this is so independently of the values of $b$ and $c$ (or
of the Markstein numbers). This  indicates that our theoretical model correctly captures the key aspects of the three hydrodynamic instabilities, which are the primary focus of this paper. 

\begin{figure}[h!]
\centering
\includegraphics[scale=0.55]{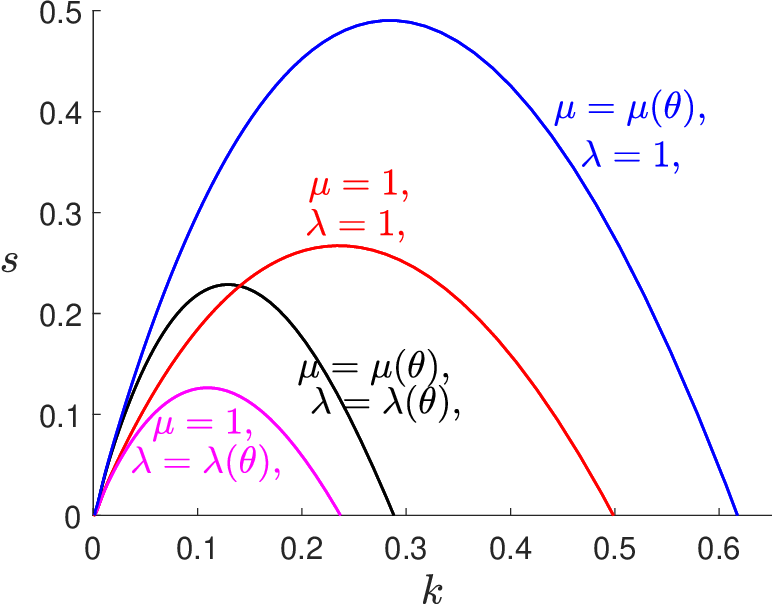}
\caption{Dispersion curves $s(k)$ computed for $\mathcal V=0$ and  $G=0$. The curves correspond to different approximations of the functions $\mu(\theta)$ and $\lambda(\theta)$.} \label{fig:disp}
\end{figure}

Before concluding this subsection, we would like to emphasise the significance of incorporating the temperature-dependent transport coefficients in the numerical dispersion relation. While $\mu(\theta)=\lambda(\theta)$ in typical reacting gaseous mixtures, the two functions play distinct roles. For instance, assuming constant viscosity ($\mu=1$) eliminates the influence of the mean flow $\mathcal V$, as the term $\mu (1- \mathcal V) \vect e_y=(1- \mathcal V) \vect e_y$ in the governing equation~\eqref{darcy1} can then be absorbed into the pressure definition. Conversely, $\lambda(\theta)$ affects the diffusive transport within the inner flame zone, as evident from~\eqref{theta1}; since $\lambda$ increases with $\theta$, accounting for this increase generally leads to larger flame speed and thickness.  Fig.~\ref{fig:disp} illustrates how the dispersion curves are modified by these two functions when  $\mathcal V=0$ and $G=0$. The black curve, computed for temperature-dependent $\mu$ and $\lambda$, provides the most accurate description, whereas the red curve, computed with $\mu=\lambda=1$, provides the least accurate description.

\subsection{Illustrative time-dependent numerical simulations}

In this subsection,   time-dependent numerical simulations  are presented to highlight the role of $\mathcal V$ and $G$ on the full development of the hydrodynamic instabilities of premixed flames. 
The computations are initiated from conditions corresponding to  the planar flame solution of~\eqref{baseflow}-\eqref{basetheta} upon which small amplitude random disturbances are superimposed. In order to keep the propagating flame within the computational domain, the coordinate system is shifted at each time step, as done in~\cite{daou2023flame,daou2023premixed,rajamanickam2024effect},
with the speed $S_T(t)$ defined by
\begin{equation}
    S_T(t) = \frac{1}{L_x}\int_{0}^{L_y} \int_{0}^{L_x} \omega\, dx dy \,
\end{equation}
where $(0,L_x)\times(0,L_y)$ is the computational domain. We note that $S_T(t)$ is a measure of the global propagation flame speed with respect to fluid 1 at $y=-\infty$ and quantifies the total instantaneous burning rate per unit transverse flame length. Given our non-dimensionalisation and initial condition corresponding to a planar flame, $S_T(0)=1$.

In the moving frame adopted, equations~\eqref{Poisson},~\eqref{theta1} and~\eqref{eqn1} are still valid provided $\vv$ is redefined to be $\vect v = \vect u + (S_T-\mathcal V)\vect e_y$, whereas equation~\eqref{darcy1} is replaced by $-\mu\vect v + \mu (S_T-\mathcal V)\vect e_y = \nabla p + (rG/m)\rho \vect e_y$. The problem is solved subject to  periodic boundary conditions in the $x$-direction along with the conditions
\begin{equation}
    \vect v=S_T(t)\,\vect e_y \,, \,\,\theta=0 \quad \text{as} \quad  y\to-\infty \quad \text{and} \quad  p=0 \,,\,\,\pfr{\theta}{y}=0 \quad \text{as} \quad  y\to +\infty  \,.
\end{equation}
The computations are carried out using COMSOL Multiphysics in a computation domain with size $L_x=100$ and $L_y=300$. The initial planar flame is positioned around $y=100$. More details on the computations can be found in~\cite{daou2023flame,daou2023premixed,rajamanickam2024effect}.

\begin{figure}[h!]
\centering 
\includegraphics[scale=0.49]{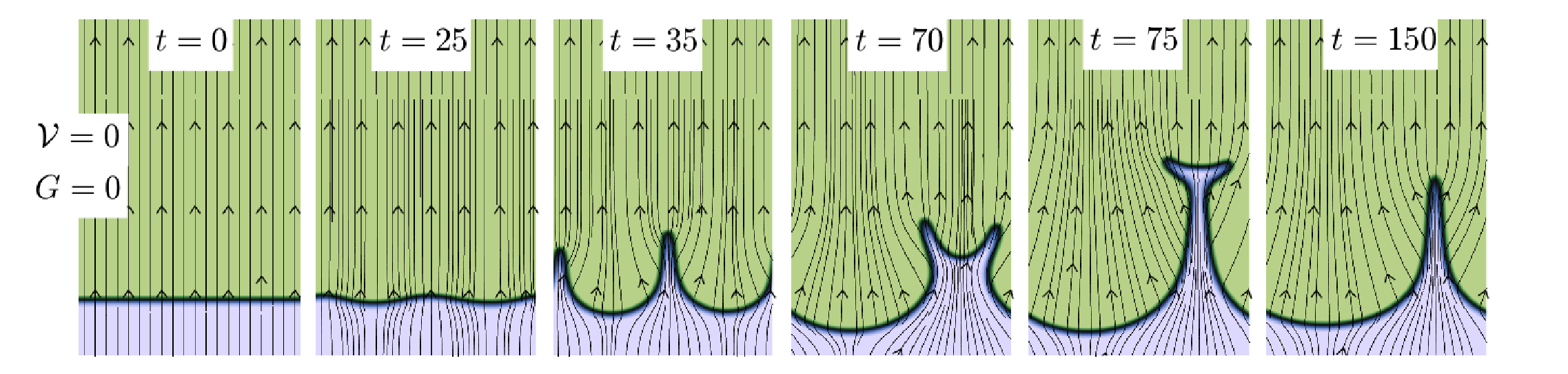} 
\includegraphics[scale=0.49]{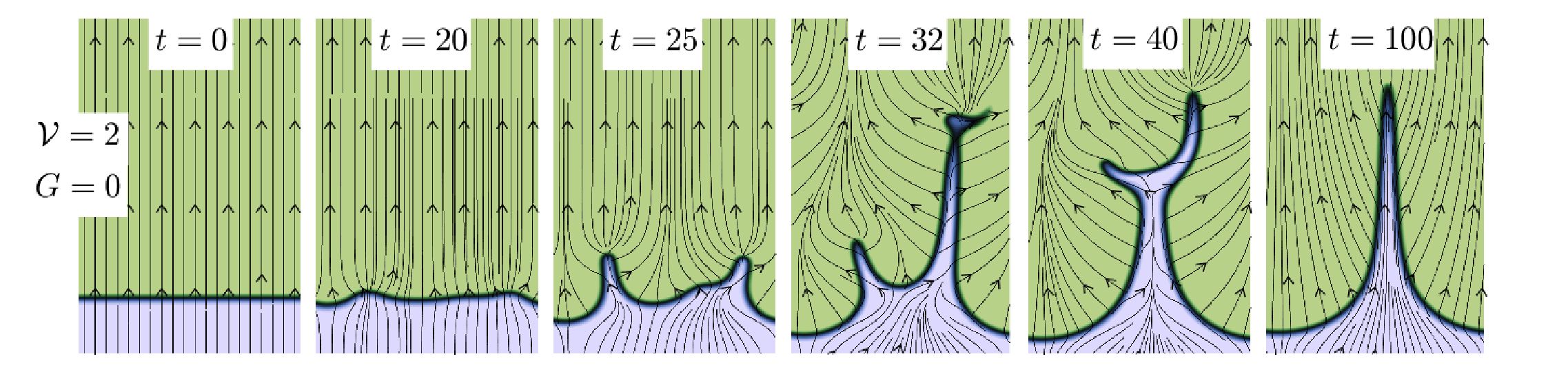} 
\includegraphics[scale=0.49]{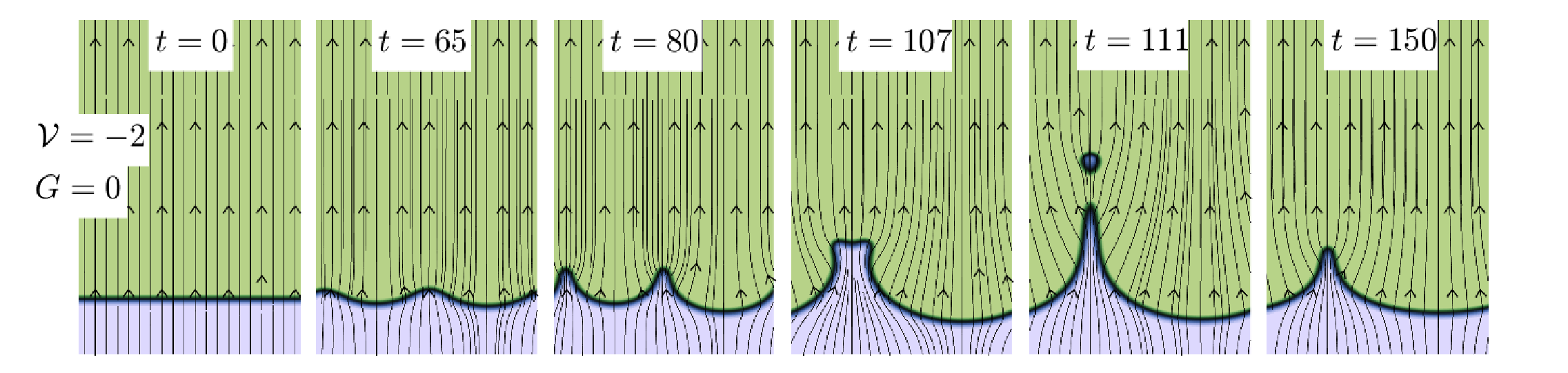} 
\caption{Instantaneous temperature fields and streamlines, with $G=0$, seen from a frame moving   with the flame speed $S_T(t)$. The upper row  of figures correspond to $\mathcal V=0$, the middle row to $\mathcal V=2$, and the lower to $\mathcal V=-2$. The horizontal non-dimensional domain size shown in each figure  is $100$ and its vertical extent $180$.} \label{fig:VG0time}
\end{figure}

\begin{figure}[h!]
\centering
\includegraphics[scale=0.55]{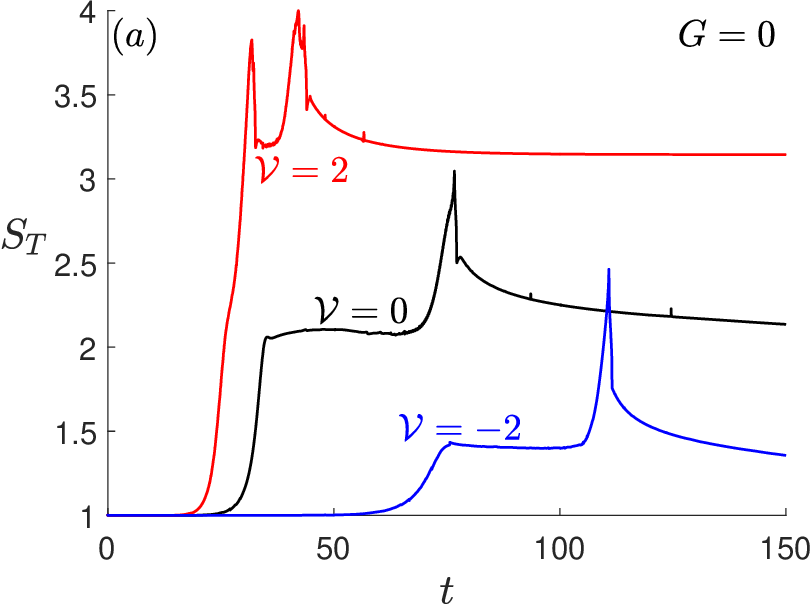}\hspace{1.5cm}
\includegraphics[scale=0.55]{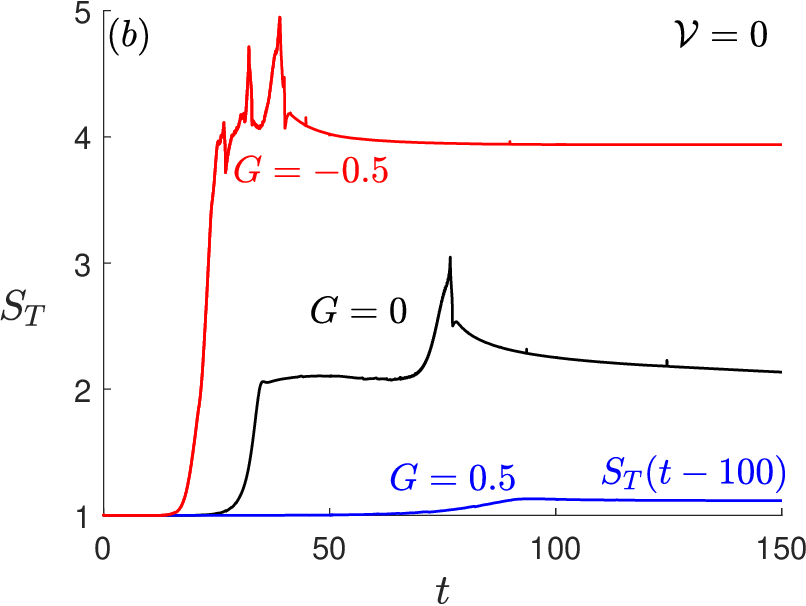}
\caption{Global propagation speed $S_T$ versus time $t$ for 
$G=0$ with selected values of $\mathcal V$ (left figure) and
$\mathcal V=0$ with selected values of $G$ (right figure).} \label{fig:ST}
\end{figure}
 
We begin by illustrating the effect of varying the imposed flow speed
$\mathcal V$ in the absence of gravity, $G=0$.  Shown in Fig.~\ref{fig:VG0time} is the temperature field $\theta(\vect x,t)$ at selected times for three cases pertaining to  $\mathcal V=0$, $2$ and $-2$. The associated global propagation speed $S_T(t)$ is plotted
in Fig.~\ref{fig:ST}(a).  For all three cases, it is observed  that the flame  develops two cusp-like structure pointing towards the burnt gas, which eventually coalesce into a single cusp settling apparently into a stable steady state. In Fig.~\ref{fig:ST}(a), the event of coalescence into a single cusp coincides with the occurrence of a peak in $S_T(t)$ before the final asymptotic behaviour. Such a behaviour is  characteristic  of the nonlinear development of Darrieus--Landau instability in flames and is qualitatively explainable using the dynamics of pole solutions of the Michelson--Sivashinsky equation~\cite{thual1988application}; the final steady solution we observe corresponds to the 1-pole solution~\cite{matalon2018darrieus}.  As for the influence of $\mathcal V$, we first note that that an increase in $\mathcal V$ has a destabilising effect, in agreement with our linear stability analysis; this can be inferred from the instability onsets observed in Fig.~\ref{fig:ST}(a).  Furthermore, it is seen that the extent of the cusp  structure (or the flame-wrinkling amplitude) increases with $\mathcal V$,  a behaviour which is typical of Saffman--Taylor instabilities known as viscous-fingering~\cite{saffman1958penetration}. 
This behaviour leads to an increase in the flame surface area with  $\mathcal V$, which 
explains the increase of the corresponding computed steady-state values of $S_T$. For a better appreciation of the full time evolution of the unstable flames, the reader is referred to the supplementary material.

\begin{figure}[h!]
\centering 
\includegraphics[scale=0.49]{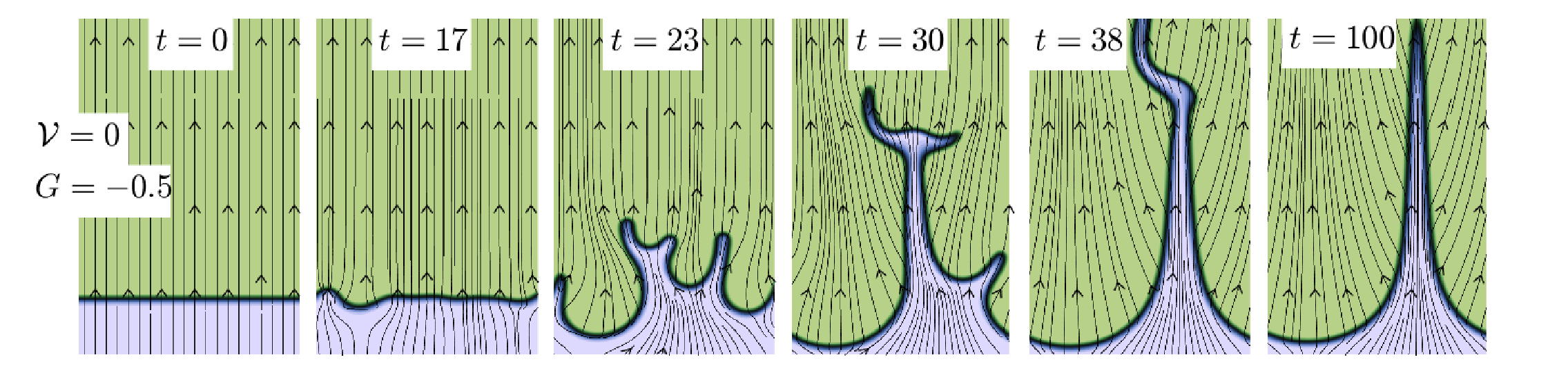} 
\includegraphics[scale=0.49]{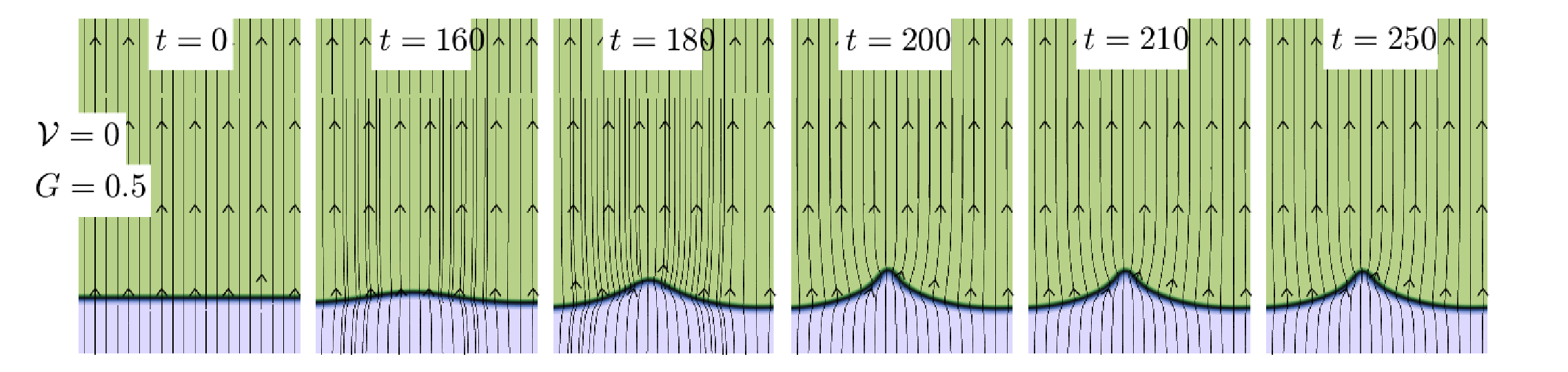} 
\caption{ Instantaneous temperature fields and streamlines, with $\mathcal V=0$, seen from a frame moving   with the flame speed $S_T(t)$. The upper row  of figures correspond to $G=-0.5$ (upward flame propagation) and the lower to $G=0.5$ (downward propagation). The horizontal non-dimensional domain size shown in each figure  is $100$ and its vertical extent $180$. } \label{fig:V0Gtime}
\end{figure}

We now briefly examine the effect of varying  the gravity parameter $G$ for a freely propagating flame, $\mathcal V=0$. The computed results are reported for two values of $G$, equal to $-0.5$ (upward flame propagation) and $0.5$ (downward flame propagation) in Fig.~\ref{fig:V0Gtime} and Fig.~\ref{fig:ST}(b) and can be compared with the gravity free case with $\mathcal V=0$ of  Fig.~\ref{fig:VG0time}. Shown in Fig.~\ref{fig:V0Gtime} is the temperature field $\theta(\vect x,t)$ at selected times and in Fig.~\ref{fig:ST}(b) the global propagation speed $S_T(t)$. The figures illustrate the expected stabilising influence of gravity for downward flame propagation, $G>0$, and destabilising influence for upward propagation. For both cases, the unstable flames settles ultimately again to  single-cusp structures albeit with markedly different wrinkling amplitudes.  We also observe that the onsets of instability for the two cases are quite distinct, with the $G=-0.5$ case being characterised by an earlier instability onset with shorter wavelength wrinkling, compared to the $G=0.5$ case. This is consistent with the linear stability analysis, in particular with the predictions of   Fig.~\ref{fig:dispVG}(b). The reader is referred to~\cite{battikh2023nonlinear} for an interesting recent study on the effect of gravity on upward propagating flames.

 \begin{figure}[h!]
\centering 
\includegraphics[scale=0.49]{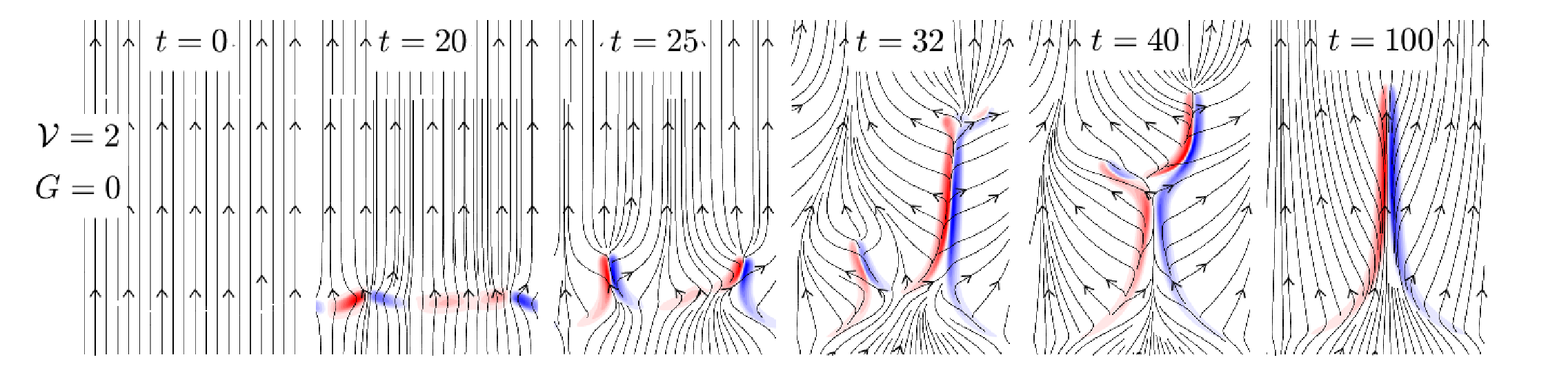} 
\caption{Instantaneous vorticity fields and streamlines with $G=0$
and $\mathcal V=2$, seen from a frame moving   with the flame speed $S_T(t)$. The horizontal non-dimensional domain size shown in each figure  is $100$ and its vertical extent $180$. The figure complements the temperature fields plotted in the middle row of Fig.~\ref{fig:VG0time}. } \label{fig:vorticity}
\end{figure}

It is insightful to examine the vorticity field, $\Omega=\nabla\times\vect v$,  in the context of Darcy's law used herein. This is obtained   by taking  the curl of the equation $\vect v -  (S_T-\mathcal V)\vect e_y = - [\nabla p + (rG/m)\rho \vect e_y]/\mu$, hence
\begin{equation}
    \boldsymbol\Omega = \Omega\vect e_z= \frac{1}{\mu^2}\nabla \mu \times \nabla p - \frac{r}{m} \nabla\left(\frac{\rho}{\mu}\right) \times G\vect e_y. 
\end{equation}
Since $\rho$ and $\mu$ vary spatially only within the interface in our model, where heat-loss and preferential diffusion are ignored, the formula indicates that the vorticity is also confined within the interface. Moreover, in the absence of gravity, $G=0$, vorticity is produced only by viscous baroclinicity associated with the misalignment of pressure and viscosity gradients, $\nabla\mu\times\nabla p \neq 0$. This implies that if viscosity is assumed constant and $G=0$, but density is allowed to vary so as to induce a Darrieus--Landau instability, then vorticity will still remain identically zero. This is in marked contrast with the usual Darrieus--Landau instability for flows governed by the Euler or Navier--Stokes equations, where vorticity is present within the flame and in the burnt gas, and vorticity-based arguments are sometimes proposed as a driving mechanism for the DL instability~\cite{matalon2018darrieus} instead of simpler arguments using mass conservation~\cite[p.354]{williams2018combustion}.  Finally, it is worth pointing out that
 Green's theorem implies
\begin{align}
   \int_{0}^{L_y} \int_{0}^{L_x}  \Omega\, dx dy &= 0 \,,
\end{align}
upon using the periodicity boundary condition in the $x$-direction. 
This explains the alternating regions of positive and negative vorticity within the flame seen in Fig.~\ref{fig:vorticity}, pertaining to the case   $\mathcal V=2$ and $G=0$ represented in the the middle row of  Fig.~\ref{fig:VG0time}. As mentioned, no vorticity is expected when viscosity is assumed constant in this case, as confirmed by computations which are not shown.

\section{Concluding remarks} \label{conc}

Propagating   interfaces in Hele-Shaw channels or porous media  are typically subject to three hydrodynamic instabilities, namely the Darrieus--Landau, Saffman--Taylor and Rayleigh--Taylor instabilities. These instabilities are long wave in nature, while at small wavelengths comparable to the interface diffusive thickness, stabilisation often occurs  due to the dependence of the interface propagation speed on its local curvature and the flow strain. Using a Darcy's flow model,
both the  long-wave  instabilities and the small-wavelength
stabilisation are captured by  a simple dispersion relation  
between the perturbation growth rate $s$ and its wavenumber $k$ of the form $s=(ak-bk^2)/(1+ck)$ given in~\eqref{nondimdisp1}. When the 
effect of the flow-strain  on the interface propagation speed  is ignored, the formula  reduces to $s=ak-bk^2$, as given in~\eqref{nondimdisp}. The parameter $a$ characterise the three hydrodynamic instabilities in a transparent way.  A key conclusion from the analysis is that a hydrodynamically unstable interface can be  fully stabilised  by controlling the imposed flow.  

In addition to these insightful results associated with  Darcy's law, a more complex dispersion relation~\eqref{nondimdispeulerdarcy}  is obtained within the so-called Euler--Darcy model, generalising those
obtained by Joulin and Sivashinsky~\cite{joulin1994influence} and
Miroshnichenko \textit{et. al.}~\cite{miroshnichenko2020hydrodynamic}.
The corresponding analysis provides a conceptual bridge between the predictions based on Darcy's law ($\varphi\to \infty$) and those based
on the Euler's equation ($\varphi\to 0$), through the use of 
a new parameter $\varphi$ defined in~\eqref{varphi}. In Hele-Shaw channels, $\varphi$ is proportional to the inverse of the channel width squared, and characterises the effect of confinement.  In the limit $\varphi \to 0$, the effect of confinement vanishes, and so does the Saffman--Taylor instability, as expected.

The theoretical results have been complemented by  a numerical stability analysis and time-dependent simulations carried out in the specific case of propagating  premixed flames in a 
Darcy flow.   The numerical results are found to be in  good agreement with the dispersion relation $s=(ak-bk^2)/(1+ck)$, in particular 
in determining the  constant $a$ which encapsulates the effects of the three hydrodynamic instabilities. Furthermore, the numerical simulations
revealed the dynamics of unstable flames under the combined influence of DL and ST instabilities, including  typical associated phenomena such as viscous fingering which significantly affects the overall burning rates.

 \begin{figure}[h!]
\centering
\includegraphics[width=0.5\textwidth]{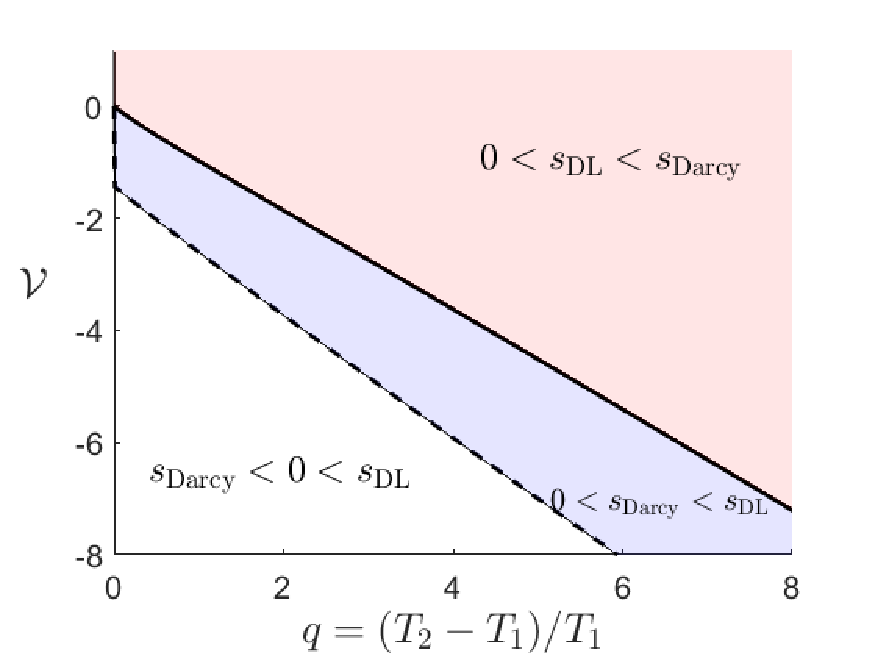}
\caption{Regions in the $q$-$\mathcal V$ plane characterising the
effect of confinement on the flame hydrodynamic instabilities  in the absence of gravity.    Note that $s_{\mathrm{DL}}$ is always positive and independent of $\mathcal V$, while $s_{\mathrm{Darcy}}$ changes with $\mathcal V$ and becomes negative in the unshaded region.} \label{fig:Vq} 
\end{figure}

We highlight now important points pertinent to the growth rates of long-wave perturbations $(k\ll 1)$  for  flames propagating in narrow channels, which are decisive in determining the conditions
for the hydrodynamic instabilities onset in  the parameter space. 
Firstly, as we have seen,  flames propagating in Hele-Shaw channels can be stabilized by an imposed flow directed from the burnt gas towards the unburnt gas $(V<0)$, and destabilised in the opposite case corresponding to  $V>0$.   Secondly, we   reemphasise   the distinction between a freely propagating flame for which $V=0$ and a flame opposed by an  imposed flow with $V=S_L^0$. The distinction is important because the  Saffman--Taylor instability which  arises when $V(\mu_2-\mu_1)\neq 0$ is absent  for a freely propagating planar flame. However, viscosity can still affect the perturbations growth rate under these conditions. For instance, in the absence of gravity and for large wavelengths ($k\ll 1$), the growth rates for the two cases   are   given by
\begin{equation*}
  \frac{s}{S_L^0 k} = \frac{\mu_2(\rho_1-\rho_2)}{\rho_2(\mu_2+\mu_1)}  \quad \text{when} \quad V=0 \quad \text{and} \quad  \frac{s}{S_L^0 k} = \frac{\rho_1\mu_2-\rho_2\mu_1}{\rho_2(\mu_2+\mu_1)}   \quad \text{when} \quad V=S_L^0 \,.
\end{equation*}
The second case ($V=S_L^0$),  originally studied by Joulin \& Sivashinsky~\cite{joulin1994influence}, is often mistakenly assumed to apply for a freely propagating flame ($V=0$) in the combustion literature.

As a final point,  it is worth comparing   the growth rate $s_{\mathrm{Darcy}}$ based on Darcy's law which is applicable under strong confinement with the growth rate $s_{\mathrm{DL}}$ corresponding to the classical Darrieus--Landau analysis which is applicable for unconfined flames. For $k\ll 1$ and in the absence of gravity, the two growth rates are given by
\begin{equation*}
    \frac{s_{\mathrm{Darcy}}}{S_L^0 k} = \frac{\rho_1\mu_2 - \rho_2 \mu_1 \mathcal V + \rho_2 \mu_2(\mathcal V-1)}{\mu_1+\mu_2} \qquad  \text{and} \qquad \frac{s_{\mathrm{DL}}}{S_L^0 k} = \frac{\rho_1}{\rho_1+\rho_2} \left(\sqrt{1 + \frac{\rho_1^2-\rho_2^2}{\rho_1\rho_2}}-1\right),
\end{equation*}
where $\mathcal V=V/S_L^0$.  The comparison of $s_{\mathrm{Darcy}}$ with $s_{\mathrm{DL}}$ produces Fig.~\ref{fig:Vq}, which has a simple yet important implication. Specifically, in the region below the solid curve, the effect of confinement (attributable to momentum loss due to friction) has a stabilising influence compared to the unconfined case,
and  a destabilising influence in the region above the curve.  Below the dashed curve, stable flames are encountered under sufficiently strong confinement.

\section*{Acknowledgements}
This research was supported by the UK EPSRC through Grants No. EP/V004840/1 and No. APP39756.

\appendix

\section{Linear stability analysis based on the Euler--Darcy model} \label{appendixa}

The main aim of this appendix is to provide    a short derivation of the dispersion relation~\eqref{dimdispeulerdarcy}   for the convenience of the reader. To this end,   to the basic solution $\left(\overline \vv,  \overline{p},\overline{f}\right)$   given by  \eqref{basicEq1}-\eqref{basicEq2} and $\overline{f}=0$, we add small perturbations  such that  $(\vv,p,f) = \left(\overline \vv,  \overline{p},\overline{f}\right) + (\vv',p',f')$.  The primed quantities satisfy on both sides of the interface the linearised form of equations~\eqref{eulerdarcymove}, namely, 
\begin{equation} \label{primedEqs}
 \nabla\cdot\vv' = 0 \,,  \quad    \rho \left(\frac{\partial \vv'}{\partial t}  + \bar v \frac{\partial \vv'}{\partial y}\right)   = -\nabla p'  - \frac {\mu}{\kappa}  \vv' \,, 
\end{equation}
from which it follows that $\nabla^2 p' =0$.  We note that \eqref{primedEqs} is a set of three independent scalar equations for the components of $\vv'=(u',v')$ and $p'$, where one of the equations may be  replaced by $\nabla^2 p' =0$, such that
\begin{equation} \label{linstabEqs}
p_{xx}'+p_{yy}'=0 \,, \quad u_x'+v_y'=0 \,, \quad 
\rho \left(v_t' + \bar v v_y'\right)= - p_y' - \frac{\mu}{\kappa} v'\,.
\end{equation}
The  equations~\eqref{linstabEqs} are subject to the interfacial conditions
\eqref{interfaceeulerdarcy} linearised at $y=0$, namely,
\begin{equation} \label{linJumpsED}
  \left \llbracket  v'  \right \rrbracket  =  \left(1-\frac{\rho_1}{\rho_2}\right) S_L^0 \mathcal{L} f'_{xx}, \,  \llbracket  p'\rrbracket =  \alpha f'  + 2 \left(\frac{\rho_1}{\rho_2}-1\right) \rho_1 {S_L^0}^2 \mathcal{L} f'_{xx}, \, \left \llbracket  u'  \right \rrbracket  =  \left(1-\frac{\rho_1}{\rho_2}\right) S_L^0  f'_{x}, \, \left. f'_t = v'\right|_{y=0^-} + S_L^0 \mathcal{L} f'_{xx}  \,,     
\end{equation}  
where $\alpha$ is given in~\eqref{alpha} and use has been made of the assumption~\eqref{SL} which implies that $S_L=S_L^0 \left (1-\mathcal{L} f'_{xx} + \cdots \right)$. The linear stability problem is therefore given by \eqref{linstabEqs},  to be solved in the domains $y>0$ and $y<0$, subject to  the interfacial conditions~\eqref{linJumpsED} at $y=0$  and the boundary conditions $p'=u'=v'= 0$   as $y \to \pm \infty$.
 
Since the problem does no depend explicitly on $x$ and $t$, we look for normal modes in the  form
\[
(f', p', u', v') =(\hat f, \hat{p}(y), \hat{u}(y),\hat{v}(y) )\exp \left( st+ikx \right) 
\]
where $s$ is in general complex, $k$  real, and $\hat f=C$  a constant.  Then, the stability problem reduces to 
\begin{equation} \label{linstabEqsHats}
\hat{p}_{yy}- k^2 \hat{p} =0 \,, \quad i k \hat u + \hat v_y=0 \,, \quad 
\rho \left(s \hat v + \bar v \hat v_y\right)= - \hat p_y - \frac{\mu}{\kappa}  \hat v \,,
\end{equation}
applicable in the domains $y>0$ and $y<0$, subject to  the interfacial conditions  
\begin{equation} \label{linJumpsEDHats}
  \left \llbracket  \hat v  \right \rrbracket  =  \left(\frac{\rho_1}{\rho_2}-1\right) S_L^0 \mathcal{L} k^2 \hat f, \,  \llbracket  \hat p \rrbracket = \left[\alpha -2 \left(\frac{\rho_1}{\rho_2}-1\right) \rho_1 {S_L^0}^2 \mathcal{L} k^2\right]\hat f, \, \left \llbracket  \hat u  \right \rrbracket  =  i \left(1-\frac{\rho_1}{\rho_2}\right) S_L^0  k \hat f, \, (s+ S_L^0 \mathcal{L} k^2) \hat f  = \hat v(0^-)      
\end{equation}  
at $y=0$, and the boundary conditions $\hat p= \hat u= \hat v= 0$   as $y \to \pm \infty$. The solution to $\hat{p}_{yy}- k^2 \hat{p} =0$ vanishing    as $y \to \pm \infty$ is given by
\[
\hat{p}= \left\{
  \begin{array}{lr}
   A e^{-k y}  &   \quad \text{for} \quad y>0 \,, \\
B e^{k y}  &    \quad \text{for} \quad y< 0 \,,
  \end{array}  
\right.
\]
where  $A$ and $B$ are constants and the wave number $k$ is assumed positive.   The solutions $\hat{u}(y)$ and  $\hat{v}(y)$ can then be determined, and these take the form 
\begin{equation}  
\hat u = \frac{i}{k} \frac{d \hat v}{dy}   \quad \text{and} \quad   \hat{v} =
     \begin{cases}\begin{aligned}
        & A\gamma_2 k e^{-k y}     + D  e^{-ky-y/\gamma_2\rho_1S_L^0}  & \text{for} \quad y > 0 \,,\\
       &-B\gamma_1 k  e^{k y}  & \text{for} \quad y < 0 \,,
    \end{aligned}\end{cases} 
\end{equation}
after enforcing the boundary conditions as $ y \to \pm \infty$, and assuming $\mathrm{Re}(s) > 0$ which is sufficient to the ultimate goal of capturing the presence of unstable modes.  Here, for brevity, we have introduced $\gamma_1 = (\rho_1 s + \mu_1/\kappa_1 + \rho_1 S_L^0 k)^{-1}$ and $\gamma_2 = (\rho_2 s + \mu_2/\kappa_2 - \rho_1 S_L^0 k)^{-1}$. Using now the  interfacial conditions~\eqref{linJumpsEDHats}, a linear homogeneous system  of  four equations for the unknown constants $A$, $B$, $C$ and $D$ is obtained, namely,
\begin{equation} \label{eq:LinStabEulerDarcySyste}
     \begin{bmatrix}
       \gamma_2k & \gamma_1k & (1-\rho_1/\rho_2) S_L^0 \mathcal{L}  k^2 & 1\\
        1 & -1 & -\alpha + 2 (\rho_1/\rho_2-1) \rho_1 {S_L^0}^2 \mathcal{L} k^2 & 0 \\
         \gamma_2k  &  -  \gamma_1k &     (1-\rho_1/\rho_2) S_L^0    k &  1+1/\rho_1S_L^0\gamma_2 k \\
        0 & \gamma_1k& s +  S_L^0 \mathcal{L} k^2& 0
    \end{bmatrix}  \begin{bmatrix}
        A   \\
        B    \\
        C \\
        D
    \end{bmatrix} =  {\bf 0}.
\end{equation} 
The solvability condition for this system, corresponding to its determinant being set to zero leads to dispersion relation~\eqref{dimdispeulerdarcy}, whose derivation is the main aim of this appendix. 

To close the appendix, we note that  the derivation of the dispersion relation~\eqref{eq:DarcyEulerDispImproved}, based on the improved propagation speed model~\eqref{SL1}, can be obtained in a similar fashion.  Without repeating the details,
the same procedure can be carried out in this case, leading ultimately to  the linear stability system  
\begin{equation}
     \begin{bmatrix}
       \gamma_2k & \gamma_1k [1+(\rho_1/\rho_2-1)   \mathcal{L}_s  k]& (1-\rho_1/\rho_2) S_L^0 \mathcal{L}_c  k^2 & 1\\
        1 & -1 -2 \gamma_1 (\rho_1/\rho_2-1) \rho_1 {S_L^0} \mathcal{L}_s k^2 & -\alpha + 2 (\rho_1/\rho_2-1) \rho_1 {S_L^0}^2 \mathcal{L}_c k^2 & 0 \\
         \gamma_2k  &  -  \gamma_1k &     (1-\rho_1/\rho_2) S_L^0    k &  1+1/\rho_1S_L^0\gamma_2 k  \\
        0 & \gamma_1k(1-\mathcal{L}_s k) & s +  S_L^0 \mathcal{L}_c k^2& 0
    \end{bmatrix}  \begin{bmatrix}
        A   \\
        B    \\
        C \\
        D
    \end{bmatrix} =  {\bf 0} \,,
\end{equation}
whose solvability  condition coincides with dispersion relation~\eqref{eq:DarcyEulerDispImproved}.

\bibliography{Combustion}

\end{document}